\documentclass[aps,prb,msmath,amssymb,amsfonts,superscriptaddress,floatfix]{revtex4} 
\pdfoutput=1
\usepackage{amsmath,amsthm,amsfonts,amssymb}
\usepackage{graphicx}

\usepackage{natbib,hyperref}
\usepackage{doi}

\usepackage{color}

\newtheorem{theorem}{Theorem}
\newtheorem{corollary}{Corollary}
\newtheorem{lemma}{Lemma}

\usepackage{algorithm}

\DeclareMathAlphabet{\mathpzc}{OT1}{pzc}{m}{it}

\newcommand{\be}{\begin{equation}}
\newcommand{\ee}{\end{equation}}

\newcommand{\pplus}{\psi_{+}}
\newcommand{\expec}{\mathbb{E}}
\newcommand{\sO}{{\mathcal O}^*}
\newcommand{\mO}{{\mathcal O}}

\newcommand{\sz}{S^{(comp)}}
\newcommand{\ent}{{\rm Ent}}
\newcommand{\Po}{P_{ov}}
\newcommand{\Ps}{P_{succ}}
\newcommand{\Pd}{P_{diab}}
\newcommand{\cfn}{\tau}
\begin{document}
\bibliographystyle{plainnat}

\title{A Short Path Quantum Algorithm for Exact Optimization}

\author{Matthew B.~Hastings}

\affiliation{Station Q, Microsoft Research, Santa Barbara, CA 93106-6105, USA}
\affiliation{Quantum Architectures and Computation Group, Microsoft Research, Redmond, WA 98052, USA}
\begin{abstract}
We give a quantum algorithm to exactly solve certain problems in combinatorial optimization, including weighted MAX-2-SAT as well as problems where the objective function is a weighted sum of products of Ising variables, all terms of the same degree $D$; this problem is called weighted MAX-E$D$-LIN2.  We require that the optimal solution be unique for odd $D$ and doubly degenerate for even $D$; however, we expect that the algorithm still works without this condition and we show how to reduce to the case without this assumption at the cost of an additional overhead.
While the time required is still exponential, the algorithm provably outperforms Grover's algorithm assuming a mild condition on the number of low energy states of the target Hamiltonian.  The detailed analysis of the runtime depends on a tradeoff between the number of such states and algorithm speed: having fewer such states allows a greater speedup.  This leads to a natural hybrid algorithm that finds either an exact or approximate solution.
\end{abstract}
\maketitle

\section{Introduction}
While quantum algorithms are useful for many problems involving linear algebra (for example, finding eigenvalues\cite{kitaev1995quantum}, solving linear equations\cite{harrow2009quantum}, and performing Hadamard transforms\cite{aaronson2015forrelation}), there are few proven speedups for combinatorial optimization problems.  The most basic such speedup is Grover's algorithm\cite{grover1996fast}, which gives a quadratic speedup over a brute-force search.  For a problem such as finding the ground state of an Ising model on $N$ spins, this can lead to a speedup from a brute force time $\sO(2^N)$ (where $\sO(\cdot)$ is big-O notation up to polylogarithmic factors, in this case polynomials in $N$) to $\sO(2^{N/2})$.  In the black box setting, Grover's algorithm is optimal\cite{bennett1997strengths}, but for problems with a structure one might try to find a further speedup.

One attempt to find a speedup is the adiabatic algorithm\cite{farhi2001quantum}.  Here, let $H_Z$ be a Hamiltonian
which is diagonal in the computational basis, such as
\be
\label{HZ}
H_Z=\sum_{i,j} J_{i,j} Z_i Z_j,
\ee
where $i$ labels different qubits and $Z_i$ is the Pauli $Z$ matrix on the $i$-th qubit.
Then, consider the Hamiltonian
\be
H_s=-(1-s)X + s H_Z,
\ee
where
\be
X=\sum_i X_i
\ee
and $X_i$ is the Pauli $X$ matrix on the $i$-th qubit.
At $s=0$, the ground state of this Hamiltonian can be easily prepared.  At $s=1$, the ground state is the ground state of $H_Z$.  If the spectral gap of $H_s$ between the ground and first excited state is only polynomially small for $s\in [0,1]$, then one may adiabatically evolve the ground state from $s=0$ to $s=1$ in polynomial time.

Unfortunately, the gap may become superpolynomially small.  Indeed, it was argued using ideas from Anderson localization\cite{altshuler2010anderson} that the gap may become as small as $N^{-{\rm const.}\times N}$ so that the time required for adiabatic evolution is of order $N^{{\rm const.}\times N}$ which is much slower than even classical brute force search.  While the actual behavior for random instances may be more complicated than this\cite{knysh2010relevance}, specific examples\cite{wecker2016training} can show this behavior.

This is perhaps not surprising.  The problem of finding the ground state of $H_Z$ is extremely hard, even if we restrict $J_{ij}\in \{-1,0,1\}$ for all $i,j$.  While some speedups are known for bounded degree (here, we consider the graph with spins as vertices and an edge between vertices if the corresponding element of $J$ is nonzero, and degree refers to the degree of this graph)\cite{furer2007exact,scott2007linear,golovnev2014new,mbh}, if we only allow polynomial space then the fastest classical algorithms for arbitrary $J$ take a time $\sO(2^{N})$.  If we allow exponential space then it is possible to reduce this time to $\sO(2^{\omega N/3})$, where $\omega$ is the matrix multiplication exponent\cite{williams2005new}.  However, not only does that algorithm require exponential space, but it is not known how to give a Grover speedup of this algorithm, so that no quantum algorithm is known taking time $\sO(2^{cN/2})$ for any $c<1$.
Finally, this algorithm is specific to constraint satisfaction problems where each constraint only involves a pair of variables, rather than a triple or more.

Here we present a quantum algorithm to find the ground state of $H_Z$ which improves on Grover's algorithm in many cases.  In the next subsection we give the problem definition and define which $H_Z$ we consider.

\subsection{Problem Definition}
The problem is to find the ground state of $H_Z$ (i.e., eigenstate of $H_Z$ with minimum eigenvalue; eigenvalues of $H_Z,H_s$ will often be called ``energies") assuming that the ground state energy $E_0$ is known.
We impose the following conditions on $H_Z$.
Let $N$ be the number of qubits.
We let $H_Z$ be any Hamiltonian that is a weighted sum of products of Pauli $Z$ operators, with each product containing exactly $D$ such operators on distinct qubits for some given $D$.  That is, the case $D=2$ is the Ising model, the case  $D=4$ is a sum of terms $Z_i Z_j Z_k Z_l$ for $i,j,k,l$ all distinct, and so on.  We take $D=\mO(1)$.  We emphasize that all products must have the same $D$ so that we do not allow for example $H_Z=Z_1 Z_2 + Z_1 Z_2 Z_3$; this is for a reason explained later.

Each product has a weight that is an integer.  We define $J_{tot}$ to be the sum of the absolute values of the weights.  We require that $J_{tot}=\mO({\rm poly}(N))$.  That is, we fix some $\beta>0$ and require $J_{tot}=\mO(N^\beta)$.  (If all weights are chosen from $\{-1,+1\}$ we have $J_{tot}=\mO(N^2)$.)

We require that $H_Z$ has a unique ground state for $D$ odd while for $D$ even we require that $H_Z$ has a doubly degenerate ground state.  For even $D$, the operator $\prod_i X_i$ commutes with the Hamiltonian, flipping all spins, so that every eigenvalue has an even degeneracy.  We call this assumption the ``degeneracy assumption".
We will analyze the performance of the algorithm without this assumption in future work.  In an Appendix we show how to reduce to the case without this assumption at the cost of an additional time overhead.

We give two theorems \ref{mainlog},\ref{mainconst} which describe the performance of the algorithm; the different theorems correspond to different choices of the parameters in the algorithm.
Both theorems show, roughly, that at least one of two things holds: the algorithm finds the ground state in a certain expected time or $H_Z$ has a large number of low energy states, i.e., eigenstates with eigenvalue close to $E_0$.

These theorems can be applied in one of two ways.  One way is to define a promise problem, in which we are promised that 
$H_Z$ does {\it not} have such a large number of low energy states as well as promised the degeneracy assumption on $H_0$.  Then, the algorithm solves this promise problem.
For this promise problem, while the algorithm is given $E_0$, it is not necessary to know $E_0$ in advance since one can try all possible $E_0$ with only polynomial overhead.

Alternatively, we can consider $H_Z$ without such a promise on the number of low energy states but still including the promise on the degeneracy assumption.  Then, we give a hybrid algorithm which tries both running the quantum algorithm here to find the exact ground state as well as random sampling (or a Grover search) to find an approximate ground state.  This hybrid algorithm will be explained after theorem \ref{mainlog}.

The problem of minimizing $H_Z$ is also known as MAX-E$D$-LIN2, as we assume that each term is exactly of degree $D$, rather than having degree at most $D$; of course, there is an overall sign difference also, as we try to minimize $H_Z$ rather than maximizing.  
Due to the uniqueness conditions on $H_Z$, we call this problem UNIQUE-MAX-E$D$-LIN2.

However, we can reduce any instance of MAX-$2$-LIN2 with a unique solution to UNIQUE-MAX-E$2$-LIN2, where MAX-$2$-LIN2 allows terms to have degree $1$ or $2$.
Up to a sign difference, MAX-$2$-LIN2 with a unique solution means that we consider $H_Z=\sum_{i,j} J_{i,j} Z_i Z_j + h_i Z_i$, with integer $J_{i,j},h_i$ and $\sum_{i,j} |J_{ij}| + \sum_i |h_i|=J_{tot}$, with $H_Z$ required to have a unique ground state.  We can find the ground state of this problem by finding one of the two ground states of a problem on $N+1$ spins defined by
$H_Z=\sum_{i,j} J_{i,j} Z_i Z_j + \sum_i h_i Z_0 Z_i$, where $0$ is the added spin.

\subsection{Main Results}
We now give the main results.
The parameters $B,K$ enter into the definition of the algorithm, given later; the algorithm makes use of a family of Hamiltonians $H_s=H_Z- s  B (X/N)^K$ so that $B$ sets the strength of a term that is proportional to a power (set by $K$) of the transverse field.
In each theorem, the statement about the number of low energy states of $H_Z$ is first expressed as a statement about probability distributions with high entropy and low energy which then implies the statement about the number of low energy states.

We give two distinct theorems.  The first theorem considers the case that
 $K=C\log(N)$ for $C>0$ a fixed constant.
 The second theorem is more general, and allows $K$ to be an arbitrary odd integer with $K\geq 3$; $K$ may depend on $N$ and the first theorem follows from the second.  The first theorem is simply given as a way to explain some of the results in the particular case of a logarithmically growing $K$, when some of the results simplify; the second theorem gives stronger results but they are more complicated.  Even in the case of a logarithmically growing $K$, the second theorem is stronger.
 
Fixed $K$, compared to logarithmically growing $K$, leads to a larger speedup but requires a stronger promise.

In an abuse of notation, when we write $K=C\log(N)$, we mean that $K$ is taken to be the smallest odd integer larger than $C \log(N)$.
We explicitly write the dependence on these constants in the equations, not hiding them in big-O notation.
The quantity $b$ is a real number, $0<b<1$.
We take $b<1$ else the second possibility (item 2 in the list) is trivially true and does not imply any interesting constraint on the density of states.

We let $W(E)$ denote the number of computational basis states with expectation value $E$ for $H_Z$.

All entropies are measured in bits rather than nats.
All logarithms are to base $2$ unless otherwise specified.


\begin{theorem}
\label{mainlog}
Assume that $H_Z$ obeys the degeneracy assumption.
Suppose that
$B=-bE_0$ and $K=C\log(N)$.
Then, at least one of the following holds:
\begin{itemize}
\item[1.] The algorithm finds the ground state in expected time
$$\sO\Bigl(2^{N/2} \exp\Bigl[-\frac{b}{2CD} \frac{N}{\log(N)} \Bigr]\Bigr).$$

\item[2.]
There is some probability distribution $p(u)$ on computational basis states with entropy at least
$$\sz\geq N\cdot (1-\mO(1)/C)$$
and with expected value of $H_Z$
at most
$(1-b) E_0+\mO(1)  \cdot \frac{J_{tot}}{N^2} C^2 D^2 \log(N)^2$.
Further, for any $\eta>0$,
for some $$E\leq E_0 + (1+\eta) \Bigl( b | E_0|+\mO(1) \cdot  \frac{J_{tot}}{N^2} C^2 D^2 \log(N)^2\Bigr),$$
we have
$$\log(W(E)) \geq N \cdot (1-\mO(1) \frac{1+\eta}{\eta}\frac{1}{C})-\frac{1+\eta}{\eta}\mO(\log(N)).$$
\end{itemize}
\end{theorem}

Theorem \ref{mainlog} implies the following corollary about a hybrid algorithm:
\begin{corollary}
There is an algorithm that, given $H_Z$ and $E_0$, either outputs 
``approximate" or ``exact".  If it outputs ``approximate", it returns also a state with
energy at most
$E_{approx}\equiv E_0 + (1.01) \Bigl( b | E_0|+\mO(1) \cdot  \frac{J_{tot}}{N^2} C^2 D^2 \log(N)^2\Bigr)$
and takes time at most
$\sO(2^{\mO(1)N/C}).$
If it outputs ``exact", it outputs an exact ground state of $H_Z$.
The expected run time of the algorithm (averaged over both approximate and exact outputs, rather than conditioned on an output) is
$\sO\Bigl(2^{N/2} \exp\Bigl[-\frac{b}{2CD} \frac{N}{\log(N)}    \Bigr]\Bigr)$.
\begin{proof}
Take $\eta=0.01$.

Run the following algorithm.  First, try repeated random sampling of states
to find a state with energy at most $E_{approx}$.
Take a total of $n_{samp}$ samples.
If any sample succeeds, terminate the algorithm, returning ``approximate" and the given state.
If item 2 of the theorem holds, each sample succeeds in finding such a state with probability
at least $2^{-\mO(1)N/C-\mO(\log(N))}=2^{-\mO(1)N/C}/{\rm poly}(N)$, so one can choose
$n_{samp}=\sO(2^{\mO(1)N/C})$ such samples and succeed with probability at
least $1-2^{-N}$.

If no sample succeeds, then run the quantum algorithm of this paper in parallel with a brute force search, until one of them finds an exact ground state, returning ``exact" and the state found.

The run time bound on the approximate output holds by construction.
We now show the expected run time bounds.
If item 1 holds, then the algorithm of this paper succeeds in expected time
$\sO\Bigl(2^{N/2} \exp\Bigl[-\frac{b}{2CD} \frac{N}{\log(N)}  \Bigr]\Bigr)$
 so the expected run time bound holds.
If item 2 holds and the repeated random sampling fails to find an approximate state then the brute force search will find the ground state in time $\sO(2^N)$; since the probability that the repeated random sampling fails is $\leq 2^{-N}$, this adds a negligible amount to the expected time of the algorithm.

Remark: the repeated random sampling can be quadratically improved with Grover search but this only improves constants which are hidden in the big-O notation.
\end{proof}
\end{corollary}

\begin{figure}
\includegraphics[width=2in]{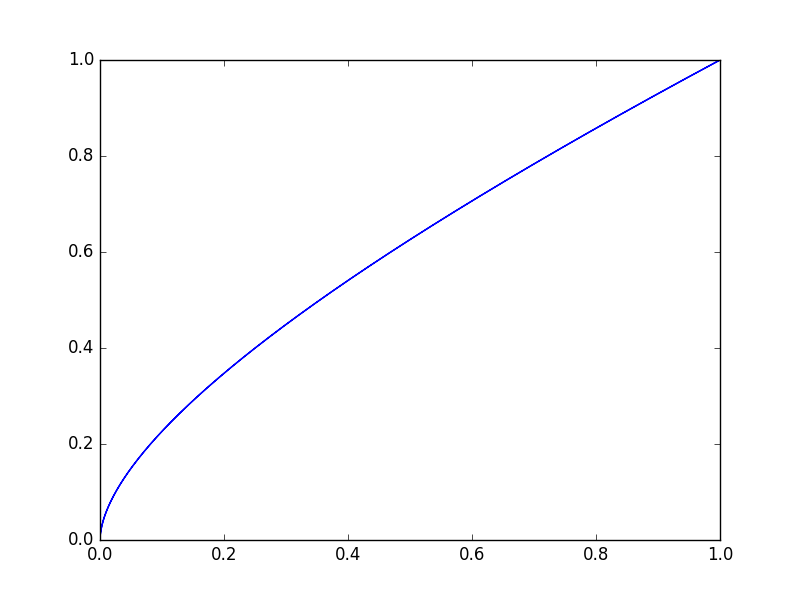}
\caption{A plot of $\cfn(\cdot)$.}
\label{cfig}
\end{figure}

We now give the case of arbitrary $K$.
The function $\cfn(\cdot)$ 
in theorem \ref{mainconst} is a continuous increasing function, taking $[0,1]$ to $[0,1]$.
It is defined in lemma \ref{lsi2}.
It is differentiable on this interval, except at zero;
for small $\sigma$,
\be
\cfn(\sigma)=\Theta(\sqrt{\frac{\sigma}{-\log(\sigma)}}).
\ee
In Fig.~\ref{cfig} we plot $\cfn(\cdot)$.

\begin{theorem}
\label{mainconst}
Assume that $H_Z$ obeys the degeneracy assumption.
Suppose
that $B=-bE_0$ with $0\leq b <1$.  Suppose that $K\geq 3$ and $K>\beta$, with $K$ odd.
Then, at least one of the following holds:
\begin{itemize}
\item[1.] The algorithm finds the ground state in expected time
$$\sO\Bigl(2^{N/2} \exp\Bigl[-\frac{b}{2DK} N       \Bigr]\Bigr).$$

\item[2.] For some $X_0\geq X_{min}=N\cdot (4B)^{-1/K}$,
there is some probability distribution $p(u)$ on computational basis states with entropy in bits at least
$$\sz\geq N \cfn^{-1}\Bigl(\frac{X_0-X_{min}/K}{N}\Bigr)$$ 
and with expected value of $H_Z$
at most
$$E_0+\mO(1) \frac{J_{tot} K^2 D^2}{X_{min}^2}+B\Bigl(\frac{X_0+X_{min}/K}{N}\Bigr)^K\cdot \mO(1).$$
Further, for some function $F(S)$ with
\be
F(S)=
E_0+\mO(1) \frac{J_{tot} K^2 D^2}{X_{min}^2}+B(\cfn(S/N))^K\cdot \mO(1),
\ee
then for some $E>E_0$ we have $\log(W(E)) \geq F^{-1}(E)-\mO(\log(N))$.
\end{itemize}
\end{theorem}
There are two parameters $b,K$ that can be adjusted in this algorithm, so the implications of this theorem take some unpacking.  We now argue, however, that for most problems for many choices of $b,K$ the conditions of items 2. of the theorem will not be satisfied and hence a nontrivial speedup will occur from item 1.  The next paragraph will deliberately be more heuristic, and will draw on some physics intuition, since the goal is not to prove a result but rather to argue that certain conditions do not occur in most examples.

Let us consider a simple example, $K=99,b=1/99$.  Then, either the algorithm
gives a speedup over a Grover search, taking expected time
$\sO(2^{cN/2})$ for a $c$ slightly less than $1$,
or there are a large number of computational basis states with low energy for $H_Z$.
Let us see whether we  can satisfy
$\log(W(E)) \geq F^{-1}(E)$ for any $E>E_0$.
For $E$ near $0$, we have $\log(W(E))$ close to $N$ since a typical state has energy $E$ near $0$.   If we had taken $b=1$, we could then satisfy $N \geq F^{-1}(0)$.  However, the function $F(S)$ is bounded at $E_0+\mO(1)\frac{J_{tot} K^2 D^2}{X_{min}^2}+ B \mO(1)$, and the constant hidden in the big-O notation $\mO(1)$ is such that for $b=1/99$ we have $F(S)$ bounded away from zero.  Indeed, $F(S)$ will be upper bounded by roughly $(1-b \mO(1)) E_0$.
Let us, for clarity in explanation pretend that the constant hidden in the $\mO(1)$ is equal to $1$.
Then, we could have a problem if $F^{-1}(\frac{98}{99} E_0) \approx N$.  However, this means an enormous (entropy almost equal to $N$) number of states have energy which is $98/99$ times the ground state energy.
Now let us instead ask whether the conditions of item 2. can be satisfied for $X_0$ close to $0$.  Indeed, for an example such as the Sherrington-Kirkpatrick spin glass\cite{kirkpatrick1975solvable}, the entropy $\log(W(E))$ has a power law dependence on $E-E_0$, with the scaling behavior well-understood\cite{crisanti2002analysis,pankov2006low} from the Parisi solution\cite{parisi1979infinite,parisi1983order} giving $\log(W(E)) \sim N\cdot ((E-E_0)/|E_0|)^{2/3}$.  By choosing $K$ large, from studying the limiting behavior of $c$ we find that we might satisfy the condition at small $E-E_0$ if $\log(W(E))\sim N\cdot ((E-E_0)/|E_0|)^{2/99}$ up to some polylogs in energy.  So, again, we would not expect to obey item 2. at small $X_0$ since $2/99<2/3$.
Note also that for $K=99$, since $B\leq \mO(N^2)$ we have $X_{min}=\Omega(N^{1-2/99})=\Omega(N^{97/99})$ so that $X_{min}$ cannot be too small.

On the other hand, a model such as the ``toy model" of Ref.~\onlinecite{wecker2016training} has a unique ground state but has $\log(W(E))\approx N/2$ for $E=E_0+1$.  Thus, for this model, we cannot consider fixed $K$ without satisfying item 2. of the theorem but we can still obtain a nontrivial speedup by taking $K=C\log(N)$, albeit not giving 
$\sO(2^{cN/2})$ with $c<1$.

Even in the case $K=C\log(N)$, theorem \ref{mainconst} gives tighter bounds than theorem \ref{mainlog}.  The entropy in theorem \ref{mainlog} is the entropy for $X_0=X_{min}$, i.e., it is the minimum possible, while the energy given in theorem \ref{mainlog} is the largest possible that can occur from theorem \ref{mainconst} for $K=C\log(N)$ and this value of energy would not occur for $X_0=X_{min}$.  Thus, theorem \ref{mainconst} gives stronger results involving a functional relationship between entropy and energy.

\subsection{Outline}
The algorithm has three distinct ideas.  First we use a short path evolution, where the initial state is not a ground state of the initial Hamiltonian.
That is, rather than trying to follow the full evolution from a transverse field Hamiltonian that is diagonal in the $X$ basis to $H_Z$ which is diagonal in the $Z$ basis, we instead follow only a ``short path", keeping a term proportional to $H_Z$ in the Hamiltonian fixed and varying a transverse field-like term from a small value to zero.  We called this term ``transverse field-like" rather than a transverse field, because we add a more complicated term $-B(X/N)^K$ that is proportional to a power of the transverse field.  This is the second idea and plays a key role in avoiding small spectral gaps.  The third idea uses measurements to obtain adiabatic evolution with exponentially small error;
similar ideas are in Ref.~\onlinecite{somma2008quantum,Poulin}.

We will show that either the algorithm succeeds in giving a nontrivial speedup or a certain spectral gap becomes small.  However, for such a gap to become small, there must be some state with a large expectation value for $-B(X/N)^K$ and with a small expectation value for $H_Z$.  Roughly, the idea then is that for a large $K$, this imposes strong constraints on the expectation value of $X$.  For example, if the state is an eigenstate of $X$, then we would have $\langle X/N \rangle = \langle (X/N)^K \rangle^{1/K}$, and so for large $K$, even a small expectation value of $(X/N)^K$ would give a large expectation value for $X/N$.  Technically, the implementation is more complicated than this because there may be large fluctuations in $X$ in the state.
Ignoring these technical details for the moment, the next idea is that given a large expectation value of $X$, we use
log-Sobolev estimates on entropy to relate the expectation value of $X$ to the entropy in the computational basis.
This is then used to show that there must be many eigenstates of $H_Z$ with low energy.

Other technical tools used include a Brillouin-Wigner perturbation theory to compute overlaps and
a method of localizing wavefunctions in $X$; this method is used to deal with the problem of large fluctuations in $X$.

In sections \ref{smpa},\ref{proj} we give the algorithm.  In section \ref{bwpt} we explain the Brillouin-Wigner perturbation theory.
We then consider two cases, depending on the 
the spectrum of the Hamiltonian $QH_s Q$, where $Q$ projects onto the states of $H_Z$ with energy greater than $E_0$. 
Theorem \ref{Qgood} in section \ref{Qgs} assumes the first case and
gives bounds on the spectral gap of $H_s$ and shows the speedup of the algorithm.  Theorem \ref{Qbad} in section \ref{Qbs} assumes the second case and shows the existence of probability distributions over computational basis states with high entropy and low energy.
Putting these results together, theorems \ref{mainlog},\ref{mainconst} follow as shown in section \ref{proof}; there, we use the high entropy and low energy probability distributions to lower bound $W(E)$.
In section \ref{discuss}, we briefly discuss some other results and also remark on a hybrid algorithm.

Section \ref{Qbs} will use the ground state degeneracy assumption for convenience, but we will show how to remove this assumption for the results of this section.  Section \ref{Qgs} will rely more heavily on the ground state degeneracy assumption; we will remove this assumption in future work using a modified perturbation theory.

\section{The Short Path Algorithm}
\label{smpa}
The  algorithm that we use is based on applying amplitude amplification\cite{brassard2002quantum} to another algorithm.
This latter algorithm is Algorithm \ref{spa}.
Remark: The case $K=1,B=cN$ corresponds to a transverse magnetic field of strength $c$. 
Also, note the direction of evolution from $s=1$ to $s=0$.

\begin{algorithm}
\caption{Short-Path (unamplified version)}
\begin{itemize}
\item[{\bf 1.}] Prepare the wavefunction in the state $\pplus=|+\rangle^{\otimes N}$.

\item[{\bf 2.}] Use the measurement algorithm of section \ref{proj} to evolve under the Hamiltonian $H_s$ from $s=1$ to $s=0$.
where
\be
\label{Hsdef}
H_s=H_Z- s  B (X/N)^K,
\ee
where
\be
\label{Tdef}
X=\sum_i X_i.
\ee
where $K$ is a positive odd integer
and $B$ is a scalar chosen later.

When we refer to a Hamiltonian $H_s$, it will be assumed that
$s \in [0,1]$ unless explicitly stated otherwise.
We pick $B>0$ so that $H_s$ has all off-diagonal entries non-positive in the computational basis.

\item[{\bf 3.}] Measure the state in the computational basis and compute the value of $H_Z$ after measuring.
If this value is equal to $E_0$ then declare success and output the computational basis state.
\end{itemize}
\label{spa}
\end{algorithm}

To analyze this algorithm for even $D$ we need a definition.
For $D$ even, we define the ``even subspace" to be the eigenspace of $\prod_i X_i$ with eigenvalue $+1$
and define the ``odd subspace" to be the eigenspace of $\prod_i X_i$ with eigenvalue $-1$.
The Hamiltonian $H_s$ commutes with $\prod_i X_i$ and $\pplus$ is in the even subspace; the measurement algorithm will be chosen to preserve the eigenvalue of $\prod_i X_i$.  When we analyze the algorithm for even $D$, all references for the remainder of the paper to the Hamiltonian will refer to the Hamiltonian projected into the even subspace and all vectors will be in the even subspace.  We will remark on this as needed.

At $s=0$, $H_{s}$ has a unique ground state; call this state $|0\rangle$; for even $D$ this state $|0\rangle$ is the unique ground state of $H_Z$ in the even subspace and it is an equal amplitude superposition of two computational basis states.  Let $\psi_{0,s}$ be the ground state of $H_s$.  By Perron-Frobenius this state is unique for $s\geq 0$ for $K$ odd; to see this, note that one can flip any given spin $K$ times to attain a single spin flip so that the matrix $H_s$ is irreducible in the computational basis, i.e., consider the term $X_i^K$ for any given $i$.
When we refer to uniqueness of a state, we mean uniqueness up to an overall phase.

We will compute the squared overlap
\be
\Po\equiv |\langle \pplus | \psi_{0,1} \rangle|^2.
\ee
later.
All state vectors that we write will be assumed to be normalized to have unit norm, except where explicitly stated otherwise later.

Suppose that the gap remains $\Omega(1/{\rm poly}(N))$ along the path.  Indeed, later we will give conditions under which the gap is $\Omega(1)$.
Then, for any $\Ps<1,\epsilon>0$
we can use the
measurement algorithm explained in section \ref{proj} to produce, with probability at least $\Po \Ps$,
a state equal to $\psi_{0,0}$ up to some error $\epsilon$.  
This algorithm takes a time that is $\mO({\rm poly}(N,(1-\Ps)^{-1}),\log(1/\epsilon))$.
This measurement algorithm has the advantage, compared to adiabatic evolution, that we can achieve a better scaling with the error $\epsilon$.

Thus, the quantum algorithm described above succeeds with probability at least $\Po\Ps-\epsilon$
 in finding the ground state of $H_Z$.
We take $\epsilon=2^{-N}$ so that the error $\epsilon$ is negligible compared to $\Po$ computed later.
We choose $\Ps=1/2$.
Hence,
applying the method of amplitude amplification to the evolution,
one obtains an algorithm which succeeds in producing the ground state of $H_Z$ in expected time $\mO(\Po^{-1/2} \Ps^{-1/2}{\rm poly}(N,\log(1/\epsilon))$.

\section{Measurement Algorithm for Adiabatic Evolution With Smaller Error}
\label{proj}
We now explain how to do step {\bf 2.} of algorithm \ref{spa}.
We explain the method in a general setting and then show that in the particular case here, it can be reduced to a single measurement.

Suppose that we have a path of Hamiltonians $H_s$, for $s\in [0,1]$.
Suppose further that all of these Hamiltonians have a unique ground state $\psi_{0,s}$, separated from the rest of the spectrum by a gap at least $\Delta=\Omega(1/{\rm poly}(N))$ and suppose that $\Vert \partial_s H_s\Vert=\mO({\rm poly}(N))$.  Suppose that for any $s$, we can simulate time evolution under $H_s$ for time $t$
up to error $\delta$ in a time that is $\mO({\rm poly}(N,t,\log(1/\delta)))$.  This holds for the Hamiltonians $H_s$ considered above so long as $B=\mO({\rm poly}(N))$ using any of several different algorithms in the literature which achieve this time\cite{QSP,LC16,BerryEtAl2014,TS,BCK15}. 

We seek a quantum algorithm that, taking a state $\psi$ as input
with overlap $\Po=|\langle \psi | \psi_{0,1}\rangle|^2$, will succeed, with probability at least $\Po\Ps$ for $\Ps$ close to $1$, in
giving an output state which is equal to $\psi_{0,0}$ up to some error $\epsilon$,
with $\epsilon$ exponentially small.  
A natural way to do this is to follow adiabatic evolution of the Hamiltonians, i.e., to simulate a time-dependent Hamiltonian which slowly changes from $H_1$ to $H_0$, taking $\psi$ as input to the evolution and $\tilde \psi$ as output.  Unfortunately, this method has two problems, one minor and one major.  
We will explain these problems before giving a different approach using measurements that avoids these problems.

The minor problem is that we must then simulate time-dependent evolution of a Hamiltonian; this problem is not too serious as for example the Taylor series approach\cite{TS} can simulate this evolution in a time that is $\mO({\rm poly}(N,t,\log(1/\delta)))$ even for slowly varying Hamiltonians.  The major problem is that there may be diabatic transitions from the ground state $\psi_{0,s}$ to some excited state along the path.  While there is some controversy about the exact error estimates in the adiabatic theorem\cite{ambainis2004elementary,reichardt2004quantum}, even the best estimates give an error that is super-polynomially small but not exponentially small in the evolution time.  Hence, even taking these estimates, in order to obtain an error $\delta$ that is exponentially small in $N$ as we will need, we need a super-polynomial evolution time.

It is possible that a careful analysis of the error terms in the adiabatic algorithm would show that the required evolution time is not in fact too large.  However, given some question about even simpler error estimates in the adiabatic theorem, we prefer not to use this method.

Instead we use the following algorithm \ref{ppa}.  Later we will show that for the specific $H_s$ considered in this paper, it suffices to use the simpler algorithm \ref{spa2} given later.  We present algorithm \ref{ppa} here because it also works for more general paths of Hamiltonians $H_s$, so long as a spectral gap is present and the adiabatic evolution can be simulated.

\begin{algorithm}
\caption{Measurement Algorithm}
\begin{itemize}
\item[{\bf 1.}] Let $\psi$ be the input state.
\item[{\bf 2.}] Phase estimate $\psi$ using $H_1$.  If the energy estimate is larger than $E_{0,1}+\Delta/2$, then
terminate the algorithm and return failure.  Else continue.
\item[{\bf 3.}] Adiabatically evolve $\psi$ from $H_1$ to $H_0$.
\item[{\bf 4.}] Phase estimate $\psi$ using $H_1$.  If the energy estimate is larger than $E_{0,1}+\Delta/2$, then
terminate the algorithm and return failure.  Else declare success and return $\psi$.
\end{itemize}
\label{ppa}
\end{algorithm}
One can also replace the adiabatic evolution of step {\bf 3.} with a sequence of measurements as in Ref.~\onlinecite{Poulin}.
This has the advantage that one does not need to simulate a time dependent evolution.

In algorithm \ref{ppa}, we use the initial phase estimation to approximately project onto $\psi_{0,1}$.  Then we adiabatically evolve.
Finally we repeat with another phase estimation to project onto $\psi_{0,0}$.
We now consider the error in the phase estimation.  We consider the phase estimation scheme of Ref.~\onlinecite{kitaev2002classical} though others are possible.
There are two ways to quantify the error.  One is the probability of error and the other is the precision.  For us, it suffices to have a precision that is smaller than $\Delta/2$ so that one distinguishes the energy of the ground and first excited states.  We take the probability of error in any step to be $\epsilon$ so that the total probability of error is $\mO(\epsilon)$.
If controlled time evolution can be implemented exactly, the phase estimation has an overhead that is only logarithmic in the error.  For constant $K$, we can use any of the time evolution algorithms above or use \cite{Poulin} to implement the controlled time evolution with error $\mO(\epsilon)$ for a time $t \sim \Delta^{-1}$ with a cost that is
$\mO({\rm poly}(\Delta^{-1},N,\log(1/\epsilon))$.

For $K=C \log(N)$, implementing the controlled time evolution is slightly more difficult.  The Hamiltonian is no longer sparse when expressed in the computational basis.  However, we can still use the approach of Ref.~\onlinecite{QSP} or the related approach in Ref.~\onlinecite{Poulin} using a sequence of measurements to implement time evolution as follows.  The term $-sB(X/N)^K$ can be expressed as a sum of terms, each of which is proportional to a projector onto an eigenspace of $X$, with eigenvalues $-N,-N+2,\ldots,N$ so that there are $N+1$ such terms.  We can measure each such term using polynomially many gates by applying a Hadamard to all qubits, use an adder to compute to total $X$, then uncomputing the addition and undoing the Hadamard.  A similar approach might also work for the Taylor series approach\cite{TS}.
Even without this way of decomposing $-sB(X/N)^K$, the overhead due to the non-sparsity of the Hamiltonian would be quasi-polynomial and would be negligible compared to the improvements in theorems \ref{mainlog},\ref{mainconst}.

We choose the adiabatic evolution so that for input state $\psi=\psi_{0,1}$, the output state has squared overlap with $\psi_{0,0}$ equal to $1-\Pd$ for some $\Pd$ close to $0$.  From the adiabatic theorems quoted above, the adiabatic evolution time required is only polynomial in $\Pd^{-1}$.  A similar estimate comes from using the repeated measurements of Ref.~\onlinecite{Poulin}.
Hence, 
\begin{lemma}
For any $\psi$, and $\Pd>0,\epsilon>0$,
the algorithm succeeds with probability at least $|\langle \psi_{0,1} | \psi \rangle|^2 (1-\Pd)-\mO(\epsilon)$.
It takes a time $\mO({\rm poly}(N,\log(1/\epsilon),\Pd^{-1})$.  Assuming success, the output state has squared overlap
at least $1-\epsilon$ with $\psi_{0,0}$.
\end{lemma}
Choosing $\Pd$ and $\epsilon$ small we obtain the desired $\Ps$ close to $1$.

In fact, for the Hamiltonian $H_s$ from section \ref{smpa}, we will show later that (under a spectral gap assumption and under some assumptions on $B,K$ that we give later that)
\be
\label{1meas}
|\langle \psi_{0,1} | \psi_{0,0} \rangle |^2=\Omega(1).
\ee
Hence,
for this problem, we can use a single measurement, phase estimating $H_{0,1}$, and not use any adiabatic evolution.
There is no need for the final phase estimation on $H_0$ since this measurement in the computational basis projects into an eigenstate of $H_Z$ whose energy can then be computed classically.

This leads to the following simple algorithm Algorithm \ref{spa2} which replaces Algorithm \ref{spa}.

\begin{algorithm}
\caption{Simplified Short-Path (unamplified version)}
\begin{itemize}
\item[{\bf 1.}] Let $\psi=\pplus$ be the input state.
\item[{\bf 2.}] Phase estimate $\psi$ using Hamiltonian $H_{1}$.
If the energy estimate is greater than $E_{0,s}+\Delta/2$, then terminate the algorithm and return failure. 
\item[{\bf 3.}] Measure the state in the computational basis and compute the value of $H_Z$ after measuring.
If this value is equal to $E_0$ then declare success and output the computational basis state.
\end{itemize}
\label{spa2}
\end{algorithm}

As a technical remark, the above algorithm assumes that we know $E_{0,1}$.  However, in the application that we consider, we do have good estimates on $E_{0,1}$, given later.

As a further remark, for even $D$ many of the Hamiltonian simulation algorithms that we refer to can be chosen to preserve the even subspace exactly.  Even if they do not preserve it exactly, they preserve it up to some error $\epsilon$ which is chosen to be negligible.  Hence, if desired, one can do the simulation using $H_s$ {\it not} projected into the even subspace.

\section{Brillouin-Wigner Perturbation Theory}
\label{bwpt}
We use Brillouin-Wigner perturbation theory to compute the ground state of $H_s$ as a function of $s$.
In the case that $H_0$ has a unique ground state (as we consider here),
Brillouin-Wigner perturbation theory gives a particularly simple result for the ground state of $H_s$, given in Eq.~(\ref{BWpt}).
There are many references for Brillouin-Wigner perturbation theory; a useful reference is Ref.~\onlinecite{leinaas1978convergence}, which also gives some convergence results.  We will however derive the needed results below.

For even $D$, we work in the even subspace throughout this section.  We remark on the appropriate choice of basis states later.

\subsection{Introduction}
We begin with some general results on Brillouin-Wigner perturbation theory.
The results in this subsection do not use any properties of the specific choice of $H_s$ above,
except that we assume that $H_0$ has a unique ground state (one great feature of this perturbation theory, however, is that it becomes only slightly more complicated when the ground state is degenerate, while Rayleigh-Schrodinger perturbation theory becomes much more complicated).
We consider a Hamiltonian $H_s=H_0+sV$ in this subsection with $H_0,V$ arbitrary, later taking $V=-B(X/N)^K$.

Let $|0\rangle$ be the ground state of $H_0$.
Let $Q=1-|0\rangle\langle 0 |$.
Let $\phi_{0,s}$ be the ground state of $H_s$.
To define the Brillouin-Wigner perturbation theory, it is convenient to normalize the states differently, rather than normalizing them to have unit norm.
We will use the normalization that
\be
\label{normalize}
\langle \phi_{0,s} | 0 \rangle=1.
\ee
so that
\be
\psi_{0,s}=\frac{\phi_{0,s}}{|\phi_{0,s}|}.
\ee
Let $E_{0,s}$ denote the energy of $\psi_{0,s}$ for Hamiltonian $H_s$.
Let
\be
G_0(\omega)=(Q (\omega-H_0) Q)^{-1},
\ee
where $\omega$ is a scalar and
where the inverse is computed in the subspace which is
the range of $Q$ and let $(1-Q)G_0=G_0(1-Q)=0$.
That is, $G_0(\omega)$ is a Moore-Penrose pseudo-inverse of $Q (\omega-H_0)) Q$, so
that $G_0(\omega) (\omega-H_0)=(\omega-H_0)G_0(\omega)=Q$.

Then Brillouin-Wigner perturbation theory gives
\begin{eqnarray}
\label{BWpt}
\phi_{0,s}&= &|0\rangle+\sum_{k\geq 1} \Bigl(sG_0(E_{0,s}) V\Bigr)^k |0\rangle.
\end{eqnarray}
Note that by definition $G_0(\omega)V=G_0(\omega) Q V$.

Note that $E_{0,s}$ appears in the power series of $\phi_{0,s}$.  Thus, in applications of this perturbation theory to compute eigenvectors or eigenvalues, it is necessary to self-consistently compute $E_{0,s}$, using
\be
\label{e0s}
E_{0,s}=E_0+s\sum_{k\geq 0}\langle 0 | V  \Bigl(sG_0(E_{0,s}) V\Bigr)^k |0\rangle
\ee
We will give bounds on $E_{0,s}$ later.

The correctness of Eq.~(\ref{BWpt}) as a formal power series in $s$ can be readily verified by computing $(H-E_{0,s}) \phi_{0,s}$.
In subsequent subsections, we will give conditions for convergence of this power series and we will compute $|\phi_{0,s}|$.
Eq.~(\ref{normalize}) is immediate from Eq.~(\ref{BWpt}).

\subsection{Overlap}
Now consider the specific choices of $H_0=H_Z$ and $V=-B(X/N)^K$ using Eq.~(\ref{Hsdef})  We will compute the overlap $\langle \pplus | \phi_{0,1}\rangle$, assuming convergence of the series (\ref{BWpt}) and assuming a bound on $E_0-E_{0,1}$.  Note that $E_{0,1} < E_0$.

For odd $D$, let us use $|u\rangle, |v\rangle,  \ldots$ to denote basis states in the computational basis.
For even $D$, we use $|u\rangle$ to denote the equal amplitude superposition of a pair of basis states in the computational basis, with the two basis states related by flipping all the spins.  Each $u$ labels a bit string of length $N$; let $\overline u$ denote the bit string with all bits flipped.  Then, $|u\rangle$ and $|\overline u\rangle$ denote the same basis state for even $D$.

Let $E_u,E_v,\ldots$ denote the corresponding eigenvalues for these states for Hamiltonian $H_0$.
Then,
\begin{eqnarray}
\label{rwseries}
\langle \pplus | \phi_{0,1} \rangle &=& \langle \pplus | 0 \rangle \\ \nonumber
&&+B\sum_{u\neq 0} \langle \pplus | u \rangle \frac{\langle u | (X/N)^K | 0 \rangle }{E_u-E_{0,1}} \\ \nonumber
&&+B^2 \sum_{u \neq 0} \sum_{v \neq 0} \langle \pplus | u \rangle \frac{\langle u | (X/N)^K | v \rangle }{E_u-E_{0,1}} \frac{\langle v | (X/N)^K | 0 \rangle}{E_v-E_{0,1}}\\ \nonumber
&&+\ldots
\end{eqnarray}
For any $u$, we have $\langle \pplus| u \rangle=2^{-N/2}$.   

Before continuing, we need two technical lemmas.  First, the following lemma which estimates
 $\langle 0 | (X/N)^L | 0 \rangle$.
\begin{lemma}
\label{XNL}
For
$0<L<N$ even, $\langle 0 | (X/N)^L | 0 \rangle \leq L!!/N^L \leq (L/N)^{L/2}$ where
$(L-1)!!=(L-1) \cdot (L-3) \cdot \ldots$.

For $L$ odd, $ \langle 0 | (X/N)^L | 0 \rangle=0$.

For $L,L'$ both even with $0<L<L'$
we have
$\langle 0 | (X/N)^L | 0 \rangle > \langle 0 | (X/N)^{L'} | 0 \rangle$, hence for
$L>N/2$ we have
$ \langle 0 | (X/N)^L | 0 \rangle \leq 2^{-N/4}$.

\begin{proof}
We have
$\langle 0 | (X/N)^L | 0 \rangle=
N^{-L} \sum_{i_1,\ldots,i_L} \langle 0 | X_{i_1} \ldots X_{i_L} | 0 \rangle$.

First consider odd $D$.
The expectation value vanishes unless for all $j$, there are an even number of $a$ such that $i_a=j$.  In that case, the expectation value is equal to $1$.  Thus, there must be some $a>1$ such that $i_a=i_1$.  There are $N$ possible choices of $i_1$ and $L-1$ possible choices of $a>1$.  Hence,
$\langle 0 | (X/N)^L | 0 \rangle \leq ((L-1)/N) \langle (X/N)^{L-2}$.  So, $\langle 0 | (X/N)^L | 0 \rangle\leq L!!/N^L\leq (L/N)^{L/2}$, where $L!!=(L-1)(L-3)\ldots$.

For even $D$, for $L\geq N$, there are additional terms in the expectation where for all $j$, there are an {\it odd} number of $a$ such that $i_a=j$.  However, since we have chosen $L<N$, such terms do not occur.

The monotonic decrease with increasing $L$ is immediate when working in the $X$ basis since $(X/N)^L > (X/N)^{L'}$.
\end{proof}
\end{lemma}

Second we need the following inequality
\begin{lemma}
\label{inverselemma}
Let $x_1,\ldots,x_k$ be positive random variables.  The variables need not be independent of each other.
Then
$\expec[\prod_{i=1}^k x_i^{-1}] \geq  \prod_{i=1}^k (\expec[x_i])^{-1}$.
\begin{proof}
We have 
$\expec[\prod_{i=1}^k x_i^{-1}]=\expec[\exp(-\sum_{i=1}^k \ln(x_i))] \geq \exp(-\expec[\sum_{i=1}^k \ln(x_i)])$, where the inequality
is by convexity of the exponential.  However, $-\expec[\ln(x_i)] \geq -\ln(\expec[x_i])$, again by convexity.
So,
$\expec[\prod_{i=1}^k x_i^{-1}] \geq \exp(-\sum_{i=1}^k \ln(\expec[x_i]))=\prod_{i=1}^k (\expec[x_i])^{-1}$.
\end{proof}
\end{lemma}

We now show that
\begin{lemma}
\label{rwsumlemma}
Assume that $E_{0,1}\geq E_0-1$ and assume that series (\ref{rwseries}) is convergent.
Let $B=\mO({\rm poly}(N))$.
Then,
\be
\label{rwsum}
\langle \pplus | \phi_{0,1} \rangle \geq 
2^{-N/2} \Bigl( \exp\Bigl[\frac{BN}{(2DK+\mO(1/N^3))|E_0|}\Bigr]-o(1) \Bigr) \frac{1}{{\rm poly}(N)}.
\ee
\begin{proof}
Note that all terms in Eq.~(\ref{rwseries}) are non-negative.  We re-express the series in terms of a random walk on the basis states $|u\rangle$ as follows.  The random walk starts in state $|0\rangle$ at time $0$.  
If the random walk is in some state $|u_t \rangle$ at time $t$, then the state of the random walk at time $t+1$ is given by repeating $K$ times the process of picking a random spin and flipping that spin.  Note that we can flip the same spin more than once in a single step of the random walk (indeed, it may be flipped up to $K$ times) although this is unlikely for $K<<\sqrt{N}$.  That is, each step of the random walk we consider here is $K$ steps of a random walk on the Boolean hypercube.

For $t>0$ we say that the random walk ``returns at time $t$" if the state of the random walk at time $t$ is $|0\rangle$.
For $t>0$ we say that the random walk ``returns by time $t$" if the random walk returns at some time $s$ with $0<s\leq t$.
Let $P_{nr}(t)$ denote the probability that the random walk does not return by time $t$.
Let $\expec_{nr,t}$ denote an expectation value conditioned on the random walk not returning by time $t$.
Then, we have
\begin{eqnarray}
\label{walk1}
\langle \pplus | \phi_{0,1} \rangle &=& 2^{-N/2} \sum_{t=0}^{\infty} B^t \expec_{nr,t}\Bigl[\prod_{m=1}^t \frac{1}{E_{u_{m}}-E_{0,1}}\Bigr] P_{nr,t}.
\end{eqnarray}
where the random walk has a sequence of states $u_1,\ldots,u_t$.

We can estimate $P_{nr,t}$ from lemma \ref{XNL}.  We have
$$P_{nr,t} \leq \sum_{0<s\leq t} \langle 0 | (X/N)^{Ks} | 0 \rangle,$$
where $\langle 0 | (X/N)^{Ks}| 0 \rangle$ is the probability that it returns at time $s$.
By lemma \ref{XNL}, for $t=\mO({\rm poly}(N))$ and $K\geq 3$, we have $P_{nr,t}=\mO(1/N^3)$.

Eq.~(\ref{walk1}) requires computing the expectation value of $\prod_{m=1}^t \frac{1}{E_{u_{m}}-E_{0,1}}$.  
Applying lemma \ref{inverselemma} to this expectation value we have
\be
\label{genconv}
\expec_{nr,t}\Bigl[\prod_{m=1}^t \frac{1}{E_{u_{m}}-E_{0,1}}\Bigr]  \geq \prod_{m=1}^t \frac{1}{\expec_{nr,t}[E_{u_m}-E_{0,1}]}.
\ee

Suppose the state of the random walk at time $t$ is given and has some energy $E_{u_t}$.  Then,
if we pick a single spin at random and flip it, the expectation value of the energy of the resulting state is equal to
$$\Bigl(1-2\frac{D}{N}\Bigr) E_{u_t}.$$
To see this, consider any term in $H_0$ which is a product of $D$ spins; the probability that we flip one of these spins, changing the sign of this term, is $D/N$.
Remark: this is the point at which we use that all terms in $H_Z$ have the same degree $D$; otherwise, the dependence of the average energy on $m$ may be more complicated.

Repeating $mK$ times,
we find that the expectation value of the energy at time $m+1$, for given $u_m$, is equal to
\begin{eqnarray}
\label{walken}
\expec[E_{u_m} ]&=&\Bigl(1-2\frac{D}{N}\Bigr)^{mK} E_{0} \\ \nonumber
&\leq & \Bigl(1-2\frac{DmK}{N}\Bigr) E_0.
\end{eqnarray}

 Since $E_{u_{m}}\geq E_0$ for all $u_m$, we have for $t=\mO({\rm poly}(N))$,
\be
\expec[E_{u_m} ] \geq (1-P_{nr,t}) E_0 + P_{nr,t} \expec_{nr,t}[E_{u_m}].
\ee
So, for $t=\mO({\rm poly}(N)$ where 
$P_{nr,t}=\mO(1/N^3)$ we have
\be
\label{walken2}
\expec_{nr,t}[E_{u_m}] \leq
\Bigl(1-2\frac{DmK}{N}-\mO(1/N^3)\Bigr) E_0.
\ee

So, by Eqs.~(\ref{genconv},\ref{walken2}) we have
\begin{eqnarray}
&&\expec_{nr,t}\Bigl[\prod_{m=1}^t \frac{1}{E_{u_{m}}-E_{0,1}}\Bigr]\\ \nonumber
&\geq &
\frac{1}{1+(\frac{2DK}{N}+\mO(1/N^3)) |E_0|}\cdot \frac{1}{1+(\frac{4DK}{N}+\mO(1/N^3)) |E_0|} \ldots \frac{1}{1+(\frac{2tDK}{N}+\mO(1/N^3)) |E_0|} \\ \nonumber
&\geq &
\Bigl(\frac{1}{(\frac{2DK}{N}+\mO(1/N^3)) |E_0|}\Bigr)^t \cdot \frac{1}{t!} \cdot \frac{1}{1+\frac{1}{(\frac{2DK}{N}+\mO(1/N^3)) |E_0|}} \cdot
\frac{1}{1+\frac{2}{(\frac{2DK}{N}+\mO(1/N^3)) |E_0|}} \cdot \ldots
\frac{1}{1+\frac{t}{(\frac{2DK}{N}+\mO(1/N^3)) |E_0|}} \\ \nonumber
&\geq &
\Bigl(\frac{1}{(\frac{2DK}{N}+\mO(1/N^3)) |E_0|}\Bigr)^t \cdot \frac{1}{t!} \cdot 
\exp\Bigl(-\frac{\log(t)+1}{1+\frac{1}{(\frac{2DK}{N}+\mO(1/N^3)) |E_0|}}\Bigr).
\end{eqnarray}
We may assume that $|E_0|=\Omega(N)$ as otherwise the ground state is degenerate.
So, for $t$ only polynomially large,
the quantity $\exp\Bigl(-\frac{\log(t)+1}{1+\frac{1}{(\frac{2DK}{N}+\mO(1/N^3)) |E_0|}}\Bigr)$
is only polynomially small; indeed for $|E_0|=\omega(N\log(N))$, this quantity is $\Omega(1)$.
So, the sum in Eq.~(\ref{rwseries}) obeys
\begin{eqnarray}
\label{limsum}
\langle \pplus | \phi_{0,1} \rangle &\geq & 2^{-N/2}  \sum_{t=0}^{\mO({\rm poly}(N))} B^t P_{nr,t} \Bigl( \frac{N}{(2 DK+\mO(1/N^3))|E_0|} \Bigr)^t \frac{1}{t!}
\frac{1}{{\rm poly}(N)}
\\ \nonumber
&=& 2^{-N/2} \Bigl( \exp\Bigl[\frac{BN}{(2DK+\mO(1/N^3))|E_0|}\Bigr]-o(1) \Bigr) \frac{1}{{\rm poly}(N)}.
\end{eqnarray}
Here, we have used the fact that 
the power series expansion of $\exp(\alpha)$ is given by $\sum_{t\geq 0} \alpha^t/t!$ and the fact that
$\sum_{t\geq s} \alpha^t/t!$ is negligible for $s>>\alpha$.  Indeed, $\sum_{t\geq s} \alpha^t/t!$ is exponentially small in $\alpha$ for fixed ratio $s/\alpha$ with $s/\alpha>e$.
Hence, we can choose the polynomial in the limits on the first line of Eq.~(\ref{limsum}) to be large compared to $\frac{BN}{(2DK+\mO(1/N^3))|E_0|}$ and then the sum of remaining terms (i.e., the terms in the series expansion of the exponential which are {\it not} included in the first line of Eq.~(\ref{limsum})) is $o(1)$.
\end{proof}
\end{lemma}

\section{Convergence Properties, Energy Shift, and Norm}
\label{Qgs}
In this subsection, we consider convergence of the series (\ref{rwseries}), we bound the shift on energy $E_0-E_{0,1}$ and we bound the norm $|\phi_{0,1}|$, and we consider the gap of $H_s$.  Most of the results will be based on considering the spectrum of $Q H_s Q$.
Again we take $H_0=H_Z$ and $V=-B(X/N)^K$.

We first show the following three results.  
Eq.~(\ref{Normgood}) below implies that Eq.~(\ref{1meas}) holds.
\begin{lemma}
\begin{itemize}
\item[1.]
The series (\ref{rwseries}) always converges, assuming that the value of $E_{0,1}$ in the series indeed is equal to the ground
state energy of $H_1$.

\item[2.] Consider the Hamiltonian $Q H_s Q$.  Let $E^{Q}_{0,s}$ be the smallest eigenvalue of this Hamiltonian in the subspace spanned by the range of $Q$.  Then, $\partial_s E^{Q}_{0,s}\leq 0$.

\item[3.] Finally, assume that $E^{Q}_{0,1}\geq E_0+1/2$.  
Then,
\be
\label{Eshift}
E_{0,1} \geq E_0 - B \langle 0 | (X/N)^K | 0 \rangle - 2 B^2 \langle 0 | (X/N)^{2K} | 0 \rangle.
\ee
and
\begin{eqnarray}
\label{Normshift}
|\phi_{0,1}|^2 & \leq &  1 + 4 B^2 \Bigl| Q(X/N)^K |0\rangle \Bigr|^2 \\ \nonumber
& \leq & 1 + 4 B^2 \langle 0 | (X/N)^{2K} | 0 \rangle.
\end{eqnarray}
For $K$ odd with $\langle 0 | B^2(X/N)^{2K}|0 \rangle\leq 1/2$, we have that
\be
\label{Egood}
E_{0,1} \geq E_0-1,
\ee
and
\be
\label{Normgood}
|\phi_{0,1}| \leq 2.
\ee
\end{itemize}
\begin{proof}
Defining
\be
G_s(\omega)=\Bigl(Q (\omega-H_s) Q\Bigr)^{-1},
\ee
where again the inverse is a Moore-Penrose inverse,
the power series (\ref{BWpt})
is a series expansion of
\be
\phi_{0,s}= |0\rangle+s G_s(E_{0,s}) V |0\rangle
\ee
in powers of $s$.
We now consider the singularities of $G_s(\omega)$ considered as a function of $s$.  
Then, the radius of convergence of the series (\ref{BWpt}) is equal to the distance from the origin $s=0$ to the closest singularity (as a function of $s$) of $G_s(\omega)$ in the complex plane.

First, we will show that
the singularities are simple poles at values of $s$ where $QH_s Q$ has 
an eigenvector in the range of $Q$ with eigenvalue equal to $\omega$.
Let $g=(Q(\omega-H_0)Q)^{-1/2}$ and let $v=QVQ$, where
the exponent $-1/2$ means a square-root of the Moore-Penrose inverse.
For $\omega$ smaller than all eigenvalues of $H_0$ (which holds for our problem since $E_{0,1}<E_0$), this inverse is well-defined and the square-root can be taken pure imaginary so that $g$ is anti-Hermitian.  We choose the sign of the square-root arbitrarily.
Then,
\be
\label{Gg}
G_s(\omega)=g(1-sgvg)^{-1}g.
\ee
To see Eq.~(\ref{Gg}), 
we have $Q(\omega-H_0 - sV) Q g (1-sgvg)^{-1} g = Q (g^{-1} -sV g) (1-sgvg)^{-1} g =
Qg^{-1} (1-sgVg) (1-sgvg)^{-1} g = Q$, where all inverses are Moore-Penrose inverses.
The matrix $gvg$ is Hermitian and so
can be diagonalized.
Thus, the singularities of $G_s(\omega)$ are simple poles at values of $s$ equal to eigenvalues of $(gvg)^{-1}$.
Note that while elsewhere in the paper we assume that $s\in[0,1]$, here we allow $s$ to be an arbitrary complex number.
Let $\zeta$ be an eigenvector of $gvg$ with eigenvalue $s^{-1}$, so that
$(1-sgvg)\zeta=0$.  Hence, $0=g (g^{-1} - svg) \zeta=g(g^{-2}-sv) g\zeta=g(\omega-H_s) g \zeta$ so that $g\zeta$
is an eigenvector of $Q H_s Q$ with eigenvalue $\omega$, as claimed.

The above results about singularities hold for arbitrary $H_0,V,\omega$.  Now to show convergence, we consider $H_0=H_Z,V=-B(X/N)^K,\omega=E_{0,1}\leq E_0$.
Suppose some matrix element of $G_s(E_{0,1})$ in the computational basis has radius of convergence $r$.   
Every matrix element of $G_s(E_{0,1})$ in the computational basis has a series expansion with all coefficients the same sign.
So, by Pringsheim's theorem, there must be a singularity for $s=r$, i.e., there must be a singularity on the positive real axis.  
So, the series for $G_1(E_{0,1})$ converges unless $E^Q_{0,t}=E_{0,1}$ for some $t\in [0,1]$.

Since all off-diagonal terms of $Q H_s Q$ and $Q V Q$ are negative, we have $\partial_s E^{Q}_{0,s}\leq 0$ by Perron-Frobenius, thus proving item 2.
 Hence, the series for $s=1$ is convergent if $E^{Q}_{0,1}>E_{0,1}$.
However, $E^{Q}_{0,1}>E_{0,1}$ also by Perron-Frobenius, thus proving item 1.

With this definition of $G_s$,
we have
\be
\label{e0swG}
E_{0,s}=E_0+s\langle 0 | V | 0\rangle +s^2 \langle 0 | V G_s(E_{0,s}) V | 0\rangle.
\ee
Let $H_0=H_Z,V=-B (X/N)^K$.
Assume that $E^{Q}_{0,1}\geq E_0+1/2$.  Since $E_{0,1}\leq E_0$,
$G_1(E_{0,1})$ has operator norm bounded by $2$.
So,
\be
E_{0,1} \geq E_0 - B \langle 0 | (X/N)^K | 0 \rangle - 2 B^2 \langle 0 | (X/N)^{2K} | 0 \rangle.
\ee

Also, if $G_1(E_{0,1})$ has operator norm bounded by $2$, we have
\begin{eqnarray}
|\phi_{0,1}|^2 &\leq &  1 + 4 B^2 \Bigl| Q(X/N)^K |0\rangle \Bigr|^2 \\ \nonumber
& \leq & 1 + 4 B^2 \langle 0 | (X/N)^{2K} | 0 \rangle.
\end{eqnarray}

This proves Eqs.~(\ref{Eshift},\ref{Normshift}).
Eqs.~(\ref{Egood},\ref{Normgood}) are immediate.
\end{proof}
\end{lemma}

We now consider the gap of $H_s$:
\begin{lemma}
The Hamiltonian $H_s$ has a gap between ground and first excited states that is greater than or equal to
$E^Q_{0,s}-E_0$.  Since  $\partial_s E^{Q}_{0,s}\leq 0$, the gap of $H_s$ is greater than or equal to $E^Q_{0,1}-E_0$.
\begin{proof}
This is a special case of a general result.  Consider a Hamiltonian
\be
\label{Hgen}
H=E_0|0\rangle\langle 0| + \sum_{a \neq 0} E_a |a\rangle\langle a|+\sum_{a\neq 0} v_a \Bigl( |a\rangle\langle 0| + {\rm h.c.} \Bigr),
\ee
where $v_a$ is an arbitrary vector.
We will show that for any $E_a,v_a,E_0$, the spectral gap of this Hamiltonian is greater than or equal to
${\rm min}_{a \neq 0}(E_a-E_0)$.
Then, to apply this result to the spectral gap of $H_s$, we take the states $|a\rangle$ in Eq.~(\ref{Hgen}) to be eigenstates of $Q H_s Q$ and take $v_a$ to be matrix elements of $sV$ between $|0\rangle$ and those eigenstates.

Define the Green's function $G(\omega)=(\omega-H)^{-1}$.
We have
\begin{eqnarray}
\langle 0 | G | 0 \rangle &=& \Bigl(\omega-E_0 -\Sigma(\omega)\Bigr)^{-1},
\end{eqnarray}
where
\be
\Sigma(\omega)=\sum_{a\neq 0}|v_a|^2 (\omega-E_a)^{-1}.
\ee

For $\omega<{\rm min}_{a \neq 0}E_a$, we have $\Sigma(\omega)<0$.
Hence, $\langle 0 | G | 0 \rangle$ does not have any poles in the interval $E_0<\omega<{\rm min}_{a \neq 0}E_a$.
Hence, if $H$ has an eigenvalue in this interval, then the corresponding eigenvector has vanishing amplitude on $|0\rangle$; however, any such eigenvector has eigenvalue equal to $E_a$ for some $a$, so no such eigenvector exists.

Thus, all eigenvalues of $H$, for any $v$ are contained in $(-\infty,E_0] \cup [{\rm min}_a(E_a),\infty)$.
If $v_a=0$ for all $a$, there is exactly one eigenvalue in the interval $(-\infty,E_0]$ and so this cannot change as $v_a$ changes.
\end{proof}
\end{lemma}

Hence it follows that:
\begin{theorem}
\label{Qgood}
Consider the Hamiltonian $Q H_s Q$.  Let $E^{Q}_{0,s}$ be the smallest eigenvalue of this Hamiltonian in the subspace spanned by the range of $Q$. Assume that $E^{Q}_{0,1}\geq E_0+1/2$ and assume that
$\langle 0 | B^2(X/N)^{2K}| 0 \rangle \leq 1/2$.
Then
\begin{itemize}
\item[1.] The Hamiltonian $H_s$ has gap at least $1/2$ between the ground and first excited state.

\item[2.] We have
\begin{eqnarray}
\langle \pplus | \psi_{0,1} \rangle & \geq &\frac{1}{2} 2^{-N/2} \Bigl( \exp\Bigl[\frac{BN}{(2DK+\mO(1/N^3))|E_0|}\Bigr]-o(1)\Bigr) \frac{1}{{\rm poly}(N)}.
\end{eqnarray}
\end{itemize}
\end{theorem}
Remark:
for $B=-bE_0$, $J_{tot}=\mO(N^\beta)$, $K\geq \beta$, the condition $\langle 0 | B^2(X/N)^{2K}| 0 \rangle \leq 1/2$ holds for all sufficiently large $N$.

\section{Gap Assumption and Entropy}
\label{Qbs}
In this section we assume that $E^Q_{0,1}<E_0+1/2$ and prove some consequences of that.
In subsection \ref{locX} we show how to construct states with large expectation value for $X$.
In subsection \ref{lSi}, we show entropic consequences of this using a log-Sobolev inequality.
In subsection \ref{conc}, we put these results together.

For technical convenience later, it is easier to work with an eigenvector of $H_1$.
So we show:
\begin{lemma}
\label{eigenvlemma}
Assume $E^Q_{0,1}<E_0+1/2$.
Then,
there is an eigenvector $\Psi$ of $H_1$ with eigenvalue at most $E_0+1/2$ such that $\langle \Psi | B(X/N)^K | \Psi \rangle \geq 1/4$.
\begin{proof}
$H_1$ has at least two eigenvalues with energy at most $E_0+1/2$.  This is because
the ground state of $Q H_1 Q$ in the subspace spanned by $Q$ and the state $|0\rangle$ span a two dimensional space.
Further, the average of energy over the corresponding eigenvectors is at most $E_0+1/4$ since the ground state energy of $H_1$ is $\leq E_0$.
On the other hand, the average of $H_Z$ over these eigenvectors is at least $E_0+1/2$.
Hence, the average of $B(X/N)^K$ over these eigenvectors is at least $1/4$.
So, at least one such $\Psi$ exists.
\end{proof}
\end{lemma}

Remark: the above lemma is the only place in this section in which the degeneracy of $H_Z$ is used.  We used the degeneracy to show that the average of $H_Z$ over these eigenvectors is at least $E_0+1/2$.
However, it is not hard to remove the degeneracy assumption here, at the cost of slightly worse constants.  We sketch this in the discussion.

\subsection{Localizing in $X$}
\label{locX}
In this section, we show how given an eigenvector with a large expectation value for $B (X/N)^K$ we can construct a wavefunction with a large expectation value for $X$ and whose expectation value for $H_Z$ is only slightly changed.
This will be necessary to apply log-Sobolev bounds in the next section. 

We begin with a lemma that considers states of a Hamiltonian describing a single particle hopping in one-dimension:
\begin{lemma}
\label{cutlemma}
Let $h$ be a real Hermitian tridiagonal matrix.  Label rows and columns of $h$ by an integer $x$, so that $h_{x,y}=0$ if $|x-y|>1$.
Let $\psi$ be a real eigenvector of $h$ with eigenvalue $E$.
Let $\psi(x)$ denote the $x$-th entry of $\psi$.  
Let $h_{od}$ denote the off-diagonal part of $h$, i.e, $h_{od}$ has the same entries as $h$ off the main diagonal and is zero on the main diagonal.
Then,
\begin{itemize}
\item[1.] For any integer $\ell>0$, for any $y$, there is a state $\xi$ with $|\xi|=1$ such that
\be
\label{cutshift}
\langle \xi | h | \xi \rangle \leq E + \mO(1/\ell^2) \Vert h_{od} \Vert,
\ee
and such that
$\xi(x)$ is non-vanishing only for $x<y+\ell$ or $\xi(x)$ is non-vanishing for $x>y-\ell$.

\item[2.]
For small $\epsilon>0$, the state $\psi_\epsilon$ defined by $\psi_\epsilon(x)=\exp(\epsilon x) \psi(x)$
obeys
\be
\label{expenergyshift}
\frac{\langle \psi_\epsilon | h | \psi_\epsilon \rangle}{|\psi_\epsilon|^2} \leq E + \mO(\epsilon^2) \Vert h_{od} \Vert.
\ee

\item[3.]
If the state $\psi_{\epsilon}$ defined above has
\be
\label{assume}
\sum_{x \geq z} \psi_\epsilon(x)^2 \geq \frac{1}{2}\sum_x \psi_\epsilon(x)^2
\ee
for some $z$,
then
there is a state
 $\xi$ with $|\xi|=1$ such that
\be
\label{cutshift2}
\langle \xi | h | \xi \rangle \leq E + \mO(1/\ell^2+\epsilon^2) \Vert h_{od} \Vert
\ee
and such that
either $\xi(x)$ is non-vanishing only for $y-\ell<x<y+\ell$ for some $y\geq z$.
\end{itemize}
\begin{proof}
Assume without loss of generality that $E=0$ (otherwise add a scalar to $h$ so that $E=0$).
Assume without loss of generality that $\psi(x)$ is real.
Let $f(x)$ be any function.  Let $\hat f$ be the diagonal matrix with entries given by $f(\cdot)$.
We compute
\begin{eqnarray}&&
\langle \hat f \psi | h | \hat f \psi \rangle\\ \nonumber
&=& \sum_x \Bigl( f(x)^2 h_{x,x} \psi(x)^2  + f(x) f(x+1) h_{x,x+1}\psi(x) \psi(x+1)+f(x) f(x-1) h_{x,x-1} \psi(x) \psi(x-1)\Bigr) \\ \nonumber
&=& \sum_x  f(x)^2 \psi(x) \Bigr( h_{x,x} \psi(x) + h_{x,x-1} \psi(x-1) + h_{x,x+1} \psi(x+1) \Bigr)\\ \nonumber
&&+\sum_x \Bigl[ f(x) (f(x+1)-f(x)) h_{x,x+1} \psi(x) \psi(x+1) + f(x) (f(x-1)-f(x)) h_{x,x-1} \psi(x) \psi(x-1) \Bigr]
\\ \nonumber
&=&\sum_x \Bigl[ f(x) (f(x+1)-f(x)) h_{x,x+1} \psi(x) \psi(x+1) + f(x) (f(x-1)-f(x)) h_{x,x-1} \psi(x) \psi(x-1) \Bigr]
\\ \nonumber
&=&\sum_x \Bigl[ f(x) (f(x+1)-f(x)) h_{x,x+1} \psi(x) \psi(x+1) + f(x+1) (f(x)-f(x+1)) h_{x+1,x} \psi(x+1) \psi(x) \Bigr]
\\ \nonumber
&=& -\sum_x (f(x)-f(x+1))^2 h_{x,x+1} \psi(x) \psi(x+1),
\end{eqnarray}
where the third equality follows from the fact that $ h_{x,x} \psi(x) + h_{x,x-1} \psi(x-1) + h_{x,x+1} \psi(x+1)=0$ since $\psi$ is an eigenvector with eigenvalue $0$ and the fourth equality follows by shifting the variable in the summation by $1$.

To prove the first claim, let $\rho^<=\sum_{y-\ell<x\leq y} \psi(x)^2$ and let $\rho^>=\sum_{y \leq x<y+\ell} \psi(x)^2$.
Assume $\rho^<>\rho^>$.
Choosing $f(x)=\ell$ for $x\leq y$ and $f(x)=\ell-(x-y)$ for $y<x<y+\ell$ and $f(x)=0$ for $x\geq y+\ell$,
we find that
$\langle \hat f \psi | h | \hat f \psi \rangle \leq \Vert h_{od} \Vert \rho^>$ and
$|\hat f\psi|^2 \geq \ell^2 \rho^<$, so that for $\xi=\hat f \psi/|\hat f \psi|$, Eq.~(\ref{cutshift}) is satisfied and $\xi(x)$ is nonvanishing only for $x<y+\ell$.  If instead $\rho^<\leq \rho^>$, choose instead
$f(x)=\ell$ for $x\geq y$ and $f(x)=\ell-|x-y|$ for $y-\ell<x<y$ and $f(x)=0$ for $x\leq y-\ell$ and 
 for $\xi=\hat f \psi/|\hat f \psi|$, Eq.~(\ref{cutshift}) is satisfied and $\xi(x)$ is nonvanishing only for $x>y-\ell$.

 To prove the second claim,
choose $f(x)=\exp(\epsilon x)$, so that we have
$\langle \hat f \psi| |h | \hat f \psi\rangle =-\sum_x \psi(x) \psi(x+1) h_{x,x+1} \mO(\epsilon^2) \exp(2\epsilon x) \leq \mO(\epsilon^2) \Vert h_{od} \Vert \cdot  |\hat f \psi|^2$.

To prove the third claim, define $\rho_y=\sum_{|x-y |\leq \ell/2} \exp(2\epsilon x) \psi(x)^2$.
If the assumption (\ref{assume}) holds, then there must be some $y>z$ such that $\rho_y\geq (1/2) \rho_{y-\ell}$; to see this, suppose no such $y$ exists, then $\rho_{z+\ell/2}< (1/2) \rho(z-\ell/2)$ and $\rho(z+3\ell/2)< (1/4)\rho(z+\ell/2)$ and so on, so $\sum_{x\geq z} \psi_\epsilon(x)^2 < \sum_{x<z} \psi_{\epsilon}(x)^2$.
 
So, there must be some $y>z$ such that $\rho_y \geq (1/2) \rho_{y-\ell}$ and such that $\rho_y\geq (1/2)\rho_{y+\ell}$.
To see this, find the largest $y>z$ such that $\rho_y \geq (1/2) \rho_{y-\ell}$; by the above paragraph at least one such $y$ must exist.  This $y$ must have
$\rho_y\geq (1/2)\rho_{y+\ell}$ as otherwise it would not be the largest.

Choose $f(x)=\exp(\epsilon x) \ell$ for $|y-x|\leq \ell$, choose
$f(x)=\exp(\epsilon x) (2\ell-|y-x|)$ for $\ell \leq |y-x| \leq 2\ell$ and choose $f(x)=0$ otherwise.
\end{proof}
\end{lemma}

The above lemma applies to a one-dimensional Hamiltonian.  However, we can apply it to
an eigenvector $\Psi$ of the Hamiltonian $H_1$ as follows.
Let $P_x$ project onto the eigenspace of $X$ with eigenvalues in the interval $[xD,xD+D)$.
That is, defining $q(y)=\lfloor y/D \rfloor$, then $P_x$ projects onto the eigenspace of $X$ with eigenvalues $y$ such
that $q(y)=x$.
Then, the Hamiltonian $H_1$, projected into the space spanned by $P_x \Psi/|P_x \Psi|$ obeys the conditions of
lemma \ref{cutlemma} with $\Vert h_{od} \Vert \leq \Vert H_Z \Vert$ and with $\psi(x)=\sqrt{\langle P_x \Psi | \Psi \rangle}$.

We now show how to attain a wavefunction with a large expectation value for $X$.  We do this in two slightly different ways depending on how large $K$ is.  The lemma will involve the constant $C>0$; the constants hidden inside the big-O notation do {\it not} depend upon $C$.
\begin{lemma}
\label{largeX}
Assume that $E^Q_{0,1}<E_0+1/2$.
\begin{itemize}
\item[1.] Assume that $K=C\log(N)$.  Let $B=-bE_0$ with $b\leq 1$.
Then, there is a state $\Xi$, with $|\Xi|=1$ such that
 \be
 \langle \Xi | X/N | \Xi \rangle\geq 1-\mO(1)/C
 \ee
 and such that 
\be
\label{e1}
\langle \Xi | H_Z | \Xi \rangle \leq 
E_0+1/2+\mO(1) \cdot  \frac{J_{tot}}{N^2} C^2 D^2 \log(N)^2+B.
\ee

\item[2.] Consider arbitrary $K$.  Let $B=-bE_0$ with $b\leq 1$.
Let $X_{min}=N\cdot (4B)^{-1/K}$.
 (Remark:
for $X<X_{min}$
we have $B(X/N)^K \leq 1/4$.)

Then, there is a state $\Xi$ with $|\Xi|=1$ such that
$\Xi$ is supported on an eigenspace of $X$ with
eigenvalues in some interval
$[X_0-X_{min}/K,X_0+X_{min}/K]$
for $X_0\geq X_{min}$ and such that
\be
\label{e2}
\langle \Xi | H_Z -B(X/N)^K| \Xi \rangle \leq E_0+1/2+\mO(1) \frac{J_{tot} K^2D^2}{X_{min}^2},
\ee
so that
\begin{eqnarray}
\label{e3}
\langle \Xi | H_Z | \Xi \rangle & \leq &E_0+1/2+\mO(1) \frac{J_{tot} K^2D^2}{X_{min}^2}+B\Bigl(\frac{X_0+X_{min}/K}{N}\Bigr)^K \\ \nonumber
 &\leq &E_0+1/2+\mO(1) \frac{J_{tot} K^2D^2}{X_{min}^2}+eB(X_0/N)^K.
 \end{eqnarray}
\end{itemize}

\begin{proof}
The proof of both cases is the same.
By lemma \ref{eigenvlemma}, there is an eigenstate $\Psi$ of $H_1$ with $\langle \Psi | H_1 |\Psi \rangle \leq E_0+1/2$ and
$\langle \Psi | B (X/N)^K | \Psi \rangle \geq 1/4$.
Throughout, the state $\psi$ will be an eigenvector of $h$ constructed from $\Psi$ as explained above
by projecting into the space spanned by $P_x \Psi/|P_x \Psi|$.

Let $X_{min}=N\cdot (4B)^{-1/K}$.
For $X<X_{min}$
we have $B(X/N)^K < 1/4$.

Recall that $K$ is chosen odd.
So
$\exp(K (X-X_{min})/X_{min})\geq 4 B (X/N)^K$,
and so
\be
\exp(K X/X_{min}) \geq 4 \exp(K) B (X/N)^K.
\ee
Hence, since $\langle \Psi | B (X/N)^K | \Psi \rangle \geq 1/4$, we have
\be
\label{bye}
\langle \Psi | \exp(K X/X_{min}) | \Psi \rangle \geq \exp(K).
\ee

Recall that $q(y)=\lfloor y/D \rfloor$.
Construct the state $\psi_{\epsilon}$ of lemma \ref{cutlemma} using $\epsilon=KD/2X_{min}$.
Then, $$|\psi_{\epsilon}|^2 \geq \langle \Psi | \exp(K Dq(X)/X_{min}) | \Psi \rangle \geq \exp(-2\epsilon)
\langle \Psi | \exp(K X/X_{min}) | \Psi\rangle \geq \exp(-2\epsilon) \exp(K),$$
using Eq.~(\ref{bye}).
At the same time,
$\sum_{x<q(X_{min}-\ln(2)X_{min}/K)} |\psi_{\epsilon}(x)|^2\leq
(1/2) \exp(K-2\epsilon)$.
So,
the state $\psi_\epsilon$ obeys
 Eq.~(\ref{assume}) for
 $z=q(X_{min}-\ln(2)X_{min}/K)$.
 Then construct state $\xi$ using
item {\bf 3} of lemma \ref{cutlemma} choosing $\ell = {\rm const.}\times \epsilon^{-1}$, using $\Vert h_{od} \Vert \leq J_{tot}$.

From state $\xi$, construct state $\Xi=\sum_x \xi(x) P_x \Psi/|P_x \Psi|$.
We have chosen $\ell={\rm const.} \times \epsilon^{-1}$; by choosing this constant smaller than $(1-\ln(2))$,
$\Xi$ will be supported on the given eigenspace of $X$.
This shows
Eq.~(\ref{e2}).  Eq.~(\ref{e1}) follows because in this case $X_{min}=N\cdot (1-\mO(1)/C)$.
\end{proof}
\end{lemma}

\subsection{Log-Sobolev Inequality}
\label{lSi}
Given a quantum state $\psi$, let $\sz(\psi)$
 be the entropy of the probability distribution of measurement outcomes when measuring the state in the computational basis.
 That is, if
 $\psi=\sum_u \psi(u) |u\rangle$, with $\sum_u |\psi(u)|^2=1$, then
 \be
 \sz(\psi)=-\sum_u |\psi(u)|^2 \log(|\psi(u)|^2).
 \ee

Here we measure entropy using bits, i.e., taking logs to base $2$, rather than nats.  This will lead to some various $\ln(2)$ differences between our definitions and definitions in the log-Sobolev literature.

In this subsection, we relate the entropy $\sz(\psi)$ to $\langle \psi | X| \psi \rangle$.  It is clear that if $\langle \psi | X| \psi \rangle=N$, then $\psi=\pplus$ up to an overall phase, and so $\sz(\psi)=N$.  Roughly, we will show that
if $\langle \psi | X | \psi \rangle$ is extensive (i.e., equal to $N$ times some nonzero constant), then $\sz(\psi)$ is also extensive.

One such result uses the log-Sobolev inequality.
This lemma \ref{lsi1} is not tight: it only gives a nontrivial bound on $\sz(\psi)$ if $\langle \psi | X | \psi \rangle > (1-\ln(2))N$.
In lemma \ref{lsi2} we give a tight bound and give a precise statement of that rough
extensivity claim above.
\begin{lemma}
\label{lsi1}
Let $\psi(u)$ be real.
Then,
\be
\label{lse}
\sz(\psi) \geq \Bigl(1-\frac{1}{\ln(2)}\Bigr) N +\frac{1}{\ln(2)} \langle \psi | X |\psi \rangle.
\ee
\begin{proof}
This follows from the log-Sobolev inequality\cite{gross1975logarithmic,o2014analysis}.
Define for any function $f(u)$,
\be
\ent(f)=\expec[f \log(\frac{f}{\expec[f]})],
\ee
where the expectation is taken for a random choice of $u$ in the domain of $f$.
We have $\expec[\psi^2]=2^{-N} \sum_u \psi(u)^2=2^{-N}$.
Here we abuse notation to use $\psi$ to represent both a quantum state and a function $\psi(u)$.
So,
\begin{eqnarray}
\label{entis}
\ent(\psi^2)&=&2^{-N} \sum_u \psi(u)^2 \Bigl( N + \log(\psi(u)^2)\Bigr)
\\ \nonumber
&=& 2^{-N}  \Bigl(N-\sz(\psi)\Bigr).
\end{eqnarray}

The log-Sobolev inequality states that
\be
\label{state}
\ent(\psi^2) \leq 2^{-N} \frac{N-\langle \psi | X | \psi \rangle}{\ln(2)}.
\ee
Hence Eq.~(\ref{lse}) follows.

Remark: the log-Sobolev literature usually expresses the inequality in terms of gradients of a function on the Boolean hypercube; however, since this gradient is defined in terms of bit flips and $X$ induces a single bit flip, after a little algebra one may see that Eq.~(\ref{state}) is equivalent to the results in the literature.

\end{proof}
\end{lemma}

A tighter bound on $\sz$ follows from the log-Sobolev inequality of Ref.~\onlinecite{samorodnitsky2008modified}.
We have
\begin{lemma}
\label{lsi2}
Let $S(x)=-x \log(x)-(1-x) \log(1-x)$ be the binary entropy function (we use $S$ rather than $H$ to avoid confusion with the Hamiltonian $H$).
Let
\be
\cfn(\sigma)=2 \sqrt{S^{-1}(\sigma) \Bigl(1-S^{-1}(\sigma)\Bigr)}\Bigr).
\ee
(The inverse of $S$ may be chosen arbitrarily so long as the same inverse is chosen in both locations.)
Then,
\be
\label{sx}
\cfn\Bigl(\frac{\sz(\psi)}{N}\Bigr)\geq  \frac{ \langle \psi | X | \psi \rangle}{N}.
\ee

The function $\cfn(\cdot)$ is a continuous increasing function, taking $[0,1]$ to $[0,1]$.
For small $\sigma$,
\be
\cfn(\sigma)=\Theta(\sqrt{\frac{\sigma}{-\log(\sigma)}}),
\ee
and for small $\langle \psi | X | \psi \rangle / N$,
\be
\label{smallscaleS}
\frac{\sz(\psi)}{N}=\Theta((X/N)^2 \log(N/X)).
\ee
\begin{proof}
This follows from theorem 1.2 of Ref.~\onlinecite{samorodnitsky2008modified}.
That theorem is expressed in terms of $\ent(\psi^2)$, up to multiplicative factors of $\ln(2)$ since natural logs are used in that paper.
Use Eq.~(\ref{entis}) to re-express this in terms of $\sz(\psi)$ and then Eq.~(\ref{sx}) follows after some algebraic manipulations.
\end{proof}
\end{lemma}

The function $\cfn(\cdot)$ has a simple interpretation.  Consider the single qubit state

$$
\sqrt{S^{-1}(\sigma)} |0\rangle +
\sqrt{1-S^{-1}(\sigma)}   | 1\rangle.$$
A product state of $N$ qubits, all in this single qubit state, has entropy $N\sigma$ and
expectation value of $X$ equal to $2N\sqrt{S^{-1}(\sigma) (1-S^{-1}(\sigma))}$.
The result is that this product state minimizes entropy for the given expectation value of $X$.  

Remark: in fact, while we have only considered the case of pure states, it is possible to show a
generalization of Eq.~(\ref{sx}) to mixed states.  We do not need this generalization here, but it is interesting to note that it exists.
Let $\rho$ be a density matrix.  Let $\sz(\rho)$ be the entropy of the mixed state obtained by measuring $\rho$ in the computational basis.  Then, we have
\be
\cfn\Bigl(\frac{\sz(\rho)}{N}\Bigr)\geq  \frac{ {\rm tr}(X \rho)}{N}.
\ee
This can be proven similarly to the proof of the pure state case: one establishes it for a single qubit, and then one uses convexity of $c$ and conditional entropy to show it for an arbitrary number of qubits.

\subsection{Number of Low Energy Eigenstates}
\label{conc}
\begin{theorem}
\label{Qbad}
Assume that $E^Q_{0,1}<E_0+1/2$.
\begin{itemize}
\item[1.] Assume that $K=C\log(N)$.  Let $B=-bE_0$ with $b\leq 1$.
Let $C=\Theta(1)$.
Then,
there is some probability distribution $p(u)$ on computational basis states with entropy at least
$$\sz\geq N\cdot (1-\mO(1)/C)$$
and with expected value of $H_Z$
at most
$(1-b)E_0+\mO(1) \cdot  \frac{J_{tot}}{N^2} C^2 D^2 \log(N)^2$.

\item[2.] Consider arbitrary $K$.  Let $B=-bE_0$ with $b\leq 1$.
Then, for some $X_0\geq X_{min}=N\cdot (4B)^{-1/K}$,
there is some probability distribution $p(u)$ on computational basis states with entropy at least
$$\sz\geq N \cfn^{-1}((X_0-X_{min}/K)/N)$$
and with expected value of $H_Z$
at most
$$E_0+\mO(1) \frac{J_{tot} K^2 D^2}{X_{min}^2}+B\Bigl(\frac{X_0+X_{min}/K}{N}\Bigr)^K\cdot \mO(1),$$
where the function $\cfn(\cdot)$ is defined in lemma \ref{lsi2}.
If $X_0<<N$, then
$$2^{N \cfn^{-1}(X_0/N)}=2^{\Omega((X_0/N)^2 \log(N/X_0)) N}$$.
\end{itemize}
\begin{proof}
This follows from lemmas
\ref{largeX},\ref{lsi1},\ref{lsi2}.
We have dropped the additive $+1/2$ from lemma \ref{largeX} as it is smaller than other terms hidden in the big-O notation.
\end{proof}
\end{theorem}

\section{Proof of Theorems \ref{mainlog},\ref{mainconst}}
\label{proof}
We now prove the main theorem.
Either $E^Q_{0,1}\geq E_0+1/2$, in which case theorem \ref{Qgood} applies or $E^Q_{0,1}<E_0+1/2$ in which case theorem \ref{Qbad} applies.
As noted, for $B=-bE_0$, for $K\geq 3$ and $K>\beta$, the condition $B(X/N)^{K}\leq 1/2$ in theorem \ref{Qgood} holds for all sufficiently large $N$.

Theorem \ref{Qbad} shows the existence of a probability distribution with large entropy and small expectation value for $H_Z$.  To turn this into a statement about
$W(E)$ as in theorems \ref{mainlog},\ref{mainconst} we need the following lemma.
Remark: the factor of $\mO(\log(N))$ in the statement of the lemma can be interpreted, for physicists, as an entropy difference arising when passing from a canonical to a microcanonical ensemble, and the need for two energies $E_1,E_2$ can be interpreted as a Maxwell construction.
\begin{lemma}
\label{entE}
Let $p(u)$ be a probability distribution over basis states $|u\rangle$ such that
\be
\sum_u p(u) E_u=\overline E.
\ee

Then, there are two energies, $E_1,E_2$ with $E_1\leq \overline E \leq E_2$
such that for some probability $P\in [0,1]$ we have $P E_1 + (1-P) E_2 = \overline E$ and $P \log(W(E_1)) + (1-P) \log(W(E_2))\geq S-\mO(\log(N))$, where $W(E)$ is the number of computational basis states with expectation value $E$ for $H_Z$.
\begin{proof}
Let
\be
P(E)=\sum_{u: E_u= E} p(u).
\ee
If $P(E)>0$, then for $u$ such that $E_u=E$, let $p(u|E)=p(u)/P(E)$ and let
\be
S(E)=-\sum_{u: E_u = E} p(u|E) \log(p(u|E)).
\ee
Then,
\be
S=-\sum_E P(E) \log(P(E))+\sum_E P(E) S(E).
\ee
Since there are only $O(J_{tot})$ possible values of $E$ for which $P(E)\neq 0$, we have $-\sum_E P(E) \log(P(E))=\mO(\log(N))$.
Hence,
$\sum_E P(E) S(E) \geq S-\mO(\log(N))$.
Also,
$\sum_E P(E) E = \overline E$.
We have $S(E)\leq \log(W(E))$.
Hence,
$\sum_E P(E) \log(W(E)) \geq S-\mO(\log(N))$.

Now maximize $\sum_E P(E) \log(W(E))$ subject to the linear constraints $\sum_E P(E) E=\overline E$ and $\sum_E P(E)=1$, with $0\leq P(E)\leq 1$ for all $E$.  Introducing Lagrange multipliers $\lambda_1,\lambda_2$ corresponding to these constraints, one finds that for all $E$ we have either $P(E)=0$ or $P(E)=1$ (and hence only one choice of $E$ has $P(E)>0$) or $\log(W(E))=\lambda_1 E + \lambda_2$.
Let $T$ be the set of $E$ such that $\log(W(E))=\lambda_1 E + \lambda_2$.
Since this constraint is linear, for any choice of $P(E)$ such that $P(E)=0$ for $E\not\in T$ and $\sum_E P(E)=1$ and 
$\sum_P(E) E=\overline E$ we have the same $\sum_E P(E) \log(W(E))$.
Hence, to maximize $\sum_E P(E) \log(W(E))$
it suffices to consider the case that $P(E)$ is nonvanishing for at most $2$ choices of $E$.  Let these choices be $E_1,E_2$ with $E_1<E_2$ and $E_1 \leq \overline E \leq E_2$.
\end{proof}
\end{lemma}
Remark: In the lemma above we used the fact that $H_Z$ has integer eigenvalues to bound the entropy $-\sum_E P(E) \log(P(E))$.  For more general $H_Z$, where $J_{ij}$ are chosen more generally, we could bin energies into polynomially many bins and obtain a similar result.

Using this lemma, now consider the first case of  theorem \ref{Qbad}, where
$K=C\log(N)$.  Theorem \ref{Qbad} shows the existence of
a probability distribution $p(u)$ on computational basis states with entropy at least
$S=\sz\geq N\cdot (1-\mO(1)/C)$
and with expected value of $H_Z$
at most
$\overline E=(1-b)E_0+1/2+\mO(1) \cdot  \frac{J_{tot}}{N^2} C^2 D^2 \log(N)^2$.
By lemma \ref{entE},
 there are two energies, $E_1,E_2$ with $E_1\leq \overline E \leq E_2$
such that for some probability $P\in [0,1]$ we have $P E_1 + (1-P) E_2 = \overline E$ and $P \log(W(E_1)) + (1-P) \log(W(E_2))=S-\mO(\log(N))$.
If $\log(W(E_1))\geq S-\mO(\log(N))$, the conclusion follows.
Otherwise, we must have $\log(W(E_2)) \geq S-\mO(\log(N))$.
If $E_2-E_0 \geq (1+\eta) (\overline E-E_0)$, then
$P_2 \leq 1/(1+\eta)$ since $E_1-E_0\geq 0$.
Since $\log(W(E_2))\leq N$, we have
$P_1 \log(W(E_1)) + P_2 N \geq S-\mO(\log(N))$ so
$\log(W(E_1)) \geq N \cdot (1-\mO(1) \frac{1+\eta}{\eta}\frac{1}{C})-\frac{1+\eta}{\eta}\mO(\log(N))$.

This completes the proof of theorem \ref{mainlog}.

Now consider the second case of theorem \ref{Qbad}, where $K$ is a constant.
The theorem shows that for some $X_0\geq X_{min}=N\cdot (4B)^{-1/K}$,
there is some probability distribution $p(u)$ on computational basis states with entropy at least
$S=\sz\geq N \cfn^{-1}((X_0-X_{min}/K)/N)$
and with expected value of $H_Z$
at most
$\overline E = E_0+\mO(1) \frac{J_{tot} K^2 D^2}{X_{min}^2}+B(\frac{X_0+X_{min}/K}{N})^K\cdot \mO(1).$
Shifting $X_0$ by $X_{min}/K$, we find that
 for some $X_0\geq (1-1/K) N\cdot (4B)^{-1/K}$,
there is some probability distribution $p(u)$ on computational basis states with entropy at least
$S=\sz\geq N \cfn^{-1}(X_0/N)$
and with expected value of $H_Z$
at most
$\overline E = E_0+\mO(1) \frac{J_{tot} K^2 D^2}{X_{min}^2}+B(\frac{X_0+2X_{min}/K}{N})^K\cdot \mO(1).$
However, for the given range of $X_0$, we have $(\frac{X_0+2X_{min}/K}{N})^K \leq (X_0/N)^K \mO(1)$, so
$\overline E \leq E_0+\mO(1) \frac{J_{tot} K^2 D^2}{X_{min}^2}+B(X_0/N)^K\cdot \mO(1).$
Now removing the restriction on the range of $X_0$ allowing $0\leq X_0 \leq N$ and defining $m_x=X_0/N$ we find that
for some $m_x\in [0,1]$ there is a probability distribution with entropy
$S\geq N \cfn^{-1}(m_x)$ and with expected value of $H_Z$ at most
\begin{eqnarray}
&&E_0+\mO(1) \frac{J_{tot} K^2 D^2}{X_{min}^2}+B(m_x)^K\cdot \mO(1)
\\ \nonumber
&\leq &E_0+\mO(1) \frac{J_{tot} K^2 D^2}{X_{min}^2}+B(\cfn(S/N))^K\cdot \mO(1) \\ \nonumber
& \equiv & F(S).
\end{eqnarray}

By lemma \ref{entE},
 there are two energies, $E_1,E_2$ with $E_1\leq \overline E \leq E_2$
such that for some probability $P\in [0,1]$ we have $P E_1 + (1-P) E_2 = \overline E\leq F(S)$ and $P \log(W(E_1)) + (1-P) \log(W(E_2))=S-\mO(\log(N))$.
The function $F(S)$ is a convex function, so either
$E_1 \leq F(\log(W(E_1))+\mO(\log(N))))$ or $E_2 \leq F(\log(W(E_2))+\mO(\log(N))))$.
This completes the proof of theorem \ref{mainconst}.

\section{Discussion}
\label{discuss}
We have given a quantum algorithm for exact optimization.
This algorithm leads to a nontrivial speedup, assuming a bound on the number of computational basis states with low
energy for $H_Z$.
This naturally leads to a hybrid algorithm: if the bound is obeyed, we run the exact algorithm here while if the bound is not obeyed, one can find a low energy state (i.e., an approximate solution of the optimization problem) by repeated random sampling or by Grover search.

We have only considered problems where all terms in $H_Z$ have the same degree.
Eq.~(\ref{walken}) is the point at which we used this assumption. For problems such as MAX-3-SAT, the energy dependence may be more complicated so it is not clear how well this algorithm works there.
It would be interesting to study this using numerical simulation.  Also we emphasize that
the mere presence of a large number of low energy states of $H_Z$ is not, on its own, sufficient to make the algorithm fail; rather, one needs to have a low energy eigenstate of $Q H_1 Q$, which requires some structure to how those states are arranged on the hypercube.  That is, if a large number of low energy states are distributed in some fairly random way on the hypercube it is possible that one might not have a small $E^Q_{0,1}$.  Numerical simulation may also be useful here.

Lemma \ref{eigenvlemma} is the only place in section \ref{Qbs} where we used the assumption on the degeneracy of $H_Z$.  However, it is not hard to remove this assumption here.
Assume $E^Q_{0,1}<E_0+1/2$.  Then consider the Hamiltonian $H_Z-(5/2) B(X/N)^K$.  This must have an eigenvalue $\leq E_0-1/4$.  Hence there must be some eigenvector $\Psi$ of $H_Z-(5/2) B(X/N)^K$ with $\langle \Psi | B(X/N)^K | \Psi \rangle \geq 1/4$.  Then one could use this $\Psi$ in the rest of section \ref{Qbs}; the cost is that in some cases factors of $B$ in this section get replaced by $(5/2) B$, in particular in theorem \ref{Qbad}.  
One can equivalently understand this as follows: one instead assumes that $E^Q_{0,2/5}<E_0+1/2$.  If this holds, then $E^Q_{0,1}<E_0-1/4$.  If this assumption fails, then section \ref{Qgood} implies a nontrivial speedup for the algorithm using Hamiltonian $H_Z-(2/5) B (X/N)^K$.
In fact, there is nothing special about our assumption that $E^Q_{0,1}<E_0+1/2$.  The analysis of section \ref{Qgs} would work under the assumption that $E^Q_{0,1}\geq E_0+c$ for any constant $c>0$.  Further, there is nothing special about our assumption in the lemma that $\langle \Psi | B(X/N)^K | \Psi \rangle \geq 1/4$; the analysis above would go through up to some changes in constants if we assumed that $\langle \Psi | B(X/N)^K | \Psi \rangle \geq c'$ for any $c'>0$.  So, it would suffice to consider $H_Z-c'' B(X/N)^K$ for $c''=(1+c')/(1-c)$ which can be made arbitrarily close to $1$.
In future work, we will address the degeneracy assumption 
in section \ref{Qgs}.

{\it Acknowledgments---} I thank J. Haah, R. Kothari, and D. Wecker for useful comments.
\bibliography{exact-ref}
\appendix
\section{Reduction to Unique Ground State}
We now corollaries how to reduce the problem without the degeneracy assumption to the problem with the degeneracy assumption, at an additional time cost.

For given $H_Z$,
consider a more general family of Hamiltonians, of the form
$H_Z+\sum_i h_i Z_i$,
where the $h_i$ are chosen from $\{-1,0,+1\}$.
We will define a sequence of such Hamiltonians, $H^{(0)}, H^{(1)}, \ldots, H^{(m)}$.  We will set $H^{(0)}=H_Z$ so that for $H^{(0)}$ all $h_i=0$.  Then, each $H^{(a+1)}$ is constructed from $H^{(a)}$ by picking some $i$ such that $h_i=0$ in the Hamiltonian $H^{(a)}$ and setting $h_i$ either equal to $+1$ or $-1$ in $H^{(a+1)}$, while leaving all other terms in the Hamiltonian unchanged.

Before continuing we need the following lemma:
\begin{lemma}
\label{halfS}
Consider any arbitrary set $S$ with $S\subseteq \{-1,+1\}^N$ and $|S|>1$.
Write elements of $S$ as vectors $v=(v_1,\ldots,v_N)$ with $v_i\in \{-1,+1\}$.
Then, there exists some $i\in \{1,\ldots,N\}$ and some $\sigma\in \{-1,+1\}$ such that
for
$$T=S \cap \{v | v_i=\sigma\}$$
we have
$1\leq |T|\leq S/2$.
In words, the set $T$ is the set of vectors in $S$ such that the $i$-th entry of the vector is equal to $\sigma$.
\begin{proof}
For each $i$, let $n_i=|S\cap \{v|v_i=+1\}|$.  If for some $i$ we have $0<n_i<|S|$, then either $1\leq n_i\leq |S|/2$ in which
case we can pick $T=S\cap \{v|v_i=+1\}$ or $S/2 \leq n_i \leq |S|-1$ in which case
we can pick $T=S\cap \{v|v_i=-1\}$.
On the other hand, if for all $i$ we have $n_i\in \{0,|S|\}$, then $|S|\leq 1$ which contradicts the assumptions of the lemma (proof: without loss of generality assume that $n_i=|S|$ for all $i$; then, only the all $+1$ vector can be in $S$).
\end{proof}
\end{lemma}

Let $H^{(a)}$ have $n_{gs}(a)$ different ground states.  Write $n_{gs}=n_{gs}(0)$.
We now 
show
\begin{lemma}
For some $m\leq \log(n_{gs})$, we can choose a sequence
 $H^{(0)},\ldots,H^{(m)}$ such that
$1 \leq n_{gs}(a+1)\leq n_{gs}(a)/2$ for all $0\leq i\leq a-1$ with
$n_{gs}(m)=1$.
\begin{proof}
The proof is inductive.  Let $S^{(a)}$ be the set of ground states of $H^{(a)}$.
We will construct each $H^{(a)}$ such that if it has $h_i\neq 0$ for any $i$, then all ground states
of $H^{(a)}$ have $v_i=-h_i$.
Then, apply lemma \ref{halfS} with $S=S^{(a)}$.  Find the $i,\sigma$ so that for
$T=S \cap \{v | v_i=\sigma\}$
we have
$1\leq |T|\leq S/2$.  By the inductive assumption $h_i=0$ in Hamiltonian $H^{(a)}$ as if $h_i\neq 0$ then
$T=\emptyset$ or $T=S^{(a)}$.  Then set $H^{(a+1)}=H^{(a)}-\sigma Z_i$.
Since the added term $-\sigma Z_i$ is equal to its minimal possible value of $-1$ on some non-empty subset of the ground states of $H^{(a)}$, those states are the ground states of $H^{(a+1)}$ and the inductive assumption holds.
Continue in this fashion, until for some $a$ we have $n_{gs}(a)=1$.
\end{proof}
\end{lemma}

Hence,
if $H_Z$ has $n_{gs}$ ground state, then there exists some Hamiltonian
$H_Z+\sum_i h_i Z_i$ with the following properties.
First, it has a unique ground state which is also a ground state of $H_Z$.
Second, the number of nonzero $h_i$ is at most $\log(n_{gs})$.
That is, writing $\vec h$ as a vector with entries $h_i$, the vector $\vec h$ has $\log(n_{gs})$ nonzero entries.

There are
$$2^1{N \choose 1} + 2^2 {N \choose 2} + \ldots + 2^{n_{gs}} {N \choose \log(n_{gs})} \leq
(2N)^{\log(n_{gs})}$$
possible choices of $\vec h$.

The Hamiltonians of the form
$H_Z+\sum_i h_i Z_i$ do not have all terms of the same degree, as some terms have degree $D$ and some have degree $1$.  However, for any such Hamiltonian
$H_Z+\sum_i h_i Z_i$ with a unique ground state we can define a Hamiltonian $H'$ on $N+D+1$ qubits which obeys the degeneracy assumption (i.e., has a unique ground state for odd $D$ or a doubly degenerate state for even $D$)
so that the ground state of $H_Z+\sum_i h_i Z_i$ can be trivially obtained from a ground state of $H'$.
To do this, add $D+1$ qubits, labelled $N+1,...,N+D+2$.
Write $J$ equal to the sum of all possible $D$-th order monomials in Pauli $Z$ operators on those added qubits, with a coefficient $-1$ in front of each monomial.  For example, for $D=3$, we have $J=-Z_{N+1} Z_{N+2} Z_{N+3}-Z_{N+1} Z_{N+2} Z_{N+4}-Z_{N+1} Z_{N+3} Z_{N+4}-Z_{N+2} Z_{N+3} Z_{N+4}$.
Consider the Hamiltonian
$$H_Z+J+\sum_{i \leq N} h_i Z_i Z_{N+1} Z_{N+2} \ldots Z_{N+D-1}.$$
The term $J$ is minimized for even $D$, by all $Z_i$ being the same for $i>N$, while for odd $D$ is it minimized by all $Z_i$ equaling $+1$ for $i>N$.
The terms $H_Z+\sum_{i \leq N} h_i Z_i Z_{N+1} Z_{N+2} \ldots Z_{N+D-1}$ can be minimized by choosing all $Z_i=+1$ and choosing the unique ground state of $H_Z+\sum_i h_i Z_i$ found above.
Thus
$H_Z+J+\sum_{i \leq N} h_i Z_i Z_{N+1} Z_{N+2} \ldots Z_{N+D-1}$ obeys the degeneracy assumption and finding its ground state directly gives a ground state of $H_Z$.

Thus, we can reduce to the unique ground state problem at the cost of trying 
$(2N)^{\log(n_{gs})}$ different choices of $\vec h$.
This number is small if $n_{gs}$ is small, while if $n_{gs}$ is large, one can find the ground state more rapidly by Grover search.  Considering the algorithm with logarithmically growing $K$, note that
if $n_{gs}\geq 2^{rN/\log(N)^2}$ for some scalar $r$ then one can find the ground state using Grover search in expected
time $\sO(2^{N/2-rN/\log(N)^2/2})$, while if
$n_{gs}\leq 2^{rN/\log(N)^2}$ then there are
only $(2N)^{rN/\log(N)^2}=2^{rN/\log(N)+\mO(N/\log(N)^2)}$ choices of $\vec h$.
For $r<b/(2CD)$, theorem \ref{mainlog} shows a speedup greater than
$2^{rN/\log(N)}$ and so one still has a $2^{{\rm const.}\times N/\log(N)}$ speedup in this case, so in worst case one has
a $2^{{\rm const.}\times N/\log(N)^2}$ speedup.
\end{document}